\documentclass[twocolumn,reprint,showpacs]{revtex4-1}
\usepackage{amsmath,amssymb}
\usepackage{dcolumn}
\usepackage{braket, bm}
\usepackage{graphicx,color}
\usepackage[colorlinks=true,citecolor=blue,linkcolor=black,urlcolor=blue]{hyperref}

\begin{document}
\title{Topological Thouless Pumping of Ultracold Fermions}
\author{Shuta Nakajima$^{1}$}\email[Electronic address: ]{shuta@scphys.kyoto-u.ac.jp} 
\author{Takafumi Tomita$^{1}$, Shintaro Taie$^{1}$, Tomohiro Ichinose$^{1}$, Hideki Ozawa$^{1}$, Lei Wang$^{2}$, Matthias Troyer$^{2}$}
\author{Yoshiro Takahashi$^{1}$}
\affiliation{$^1$\mbox{Department of Physics, Graduate School of Science, Kyoto University, Japan 606-8502}\\
$^2$\mbox{Theoretische Physik, ETH Zurich, 8093 Zurich, Switzerland}}



\begin{abstract}
A gas of electrons in a one-dimensional periodic potential can be transported even in the absence of a voltage bias if the potential is modulated slowly and periodically in time. Remarkably, the transferred charge per cycle is only sensitive to the topology of the path in  parameter space. Although this so-called Thouless charge pump has first been proposed  more than thirty years ago~\cite{Thouless}, it has not yet been realized. Here we report the first demonstration of topological Thouless pumping using ultracold atoms in a dynamically controlled optical superlattice.
We observe a shift of the atomic cloud as a result of pumping and extract the topological invariance of the pumping process from this shift. We demonstrate the topological nature of the Thouless pump by varying the topology of the pumping path and verify that the topological pump indeed works in the quantum region by varying speed and temperature.
\end{abstract}
\maketitle


Topology manifests itself in physics in a variety of ways~\cite{RevModPhys.51.591, thouless1998topological, RevModPhys.82.3045}, 
with the integer quantum Hall effect (IQHE) being one of the best-known examples in condensed matter systems. There, the Hall conductance of a two-dimensional electron gas is quantized very precisely in units of fundamental constants~\cite{PhysRevLett.45.494}. As discussed in the celebrated Thouless-Kohmoto-Nightingale-den Nijs paper~\cite{TKNN}, 
this quantized value is given by a topological invariant, the sum of the Chern numbers of the occupied energy bands.

In 1983, Thouless considered a seemingly different phenomenon of quantum transport of an electron gas in an infinite one-dimensional periodic potential, driven in a periodic cycle~\cite{Thouless}. This appears to be similar to the famous Archimedes screw~\cite{Altshuler},
which pumps water via a rotating spiral tube. However, while the Archimedes screw follows classical physics and the  pumped amount of water can be continuously changed by tilting the screw, the charge pumped by the Thouless pump is a topological quantum number and not affected by a smooth change of parameters~\cite{Thouless}.
Interestingly, this quantization of pumped charge shares the same topological origin as the IQHE. The  charge pumped per cycle can be expressed by the Chern number defined over a (1+1) dimensional periodic  Brillouin zone formed by quasimomentum $k$ and time $t$.
Although several single electron pumping experiments have been implemented in nanoscale devices 
such as quantum dots with modulated gate voltages~\cite{Switkes,Blumenthal,Kaestner}
or surface acoustic waves in order to create a potential periodic in time~\cite{Shilton},
the topological Thouless pump, which should have the spatial periodicity to define the Bloch wave function as well as the temporal periodicity, has not been realized in electron systems.  

\begin{figure}[tb]
\includegraphics[width=8cm,clip]{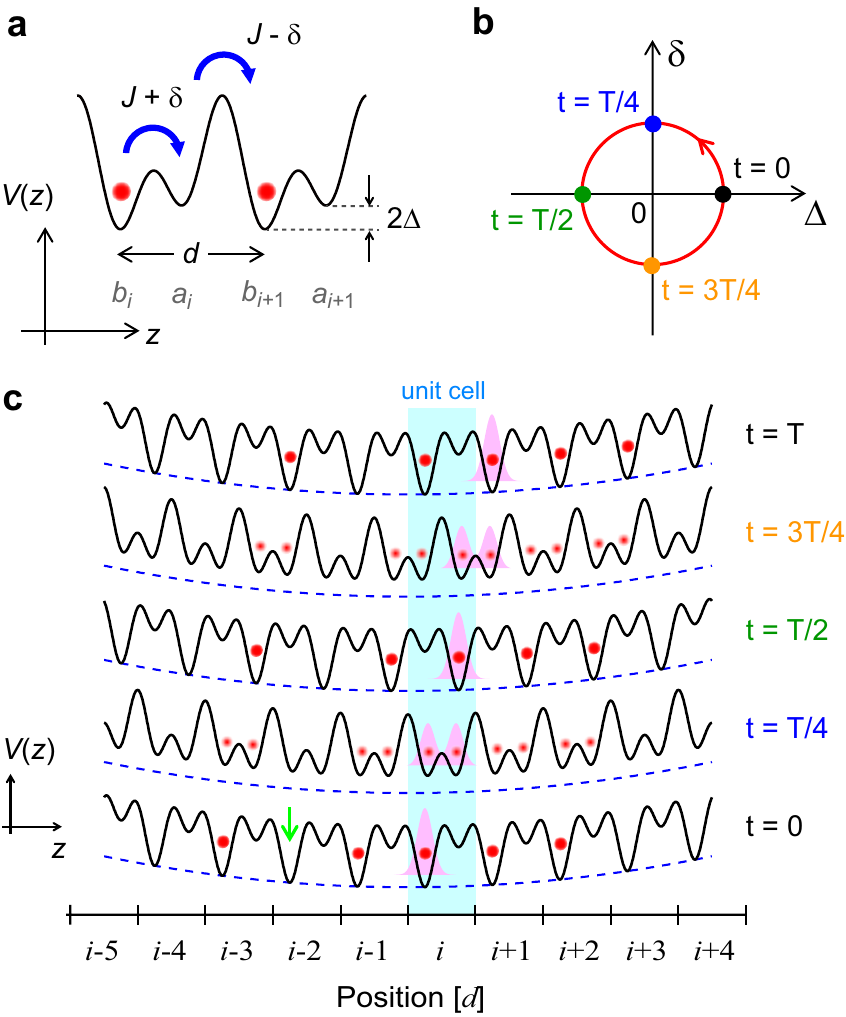}
\caption{
{\bf The Rice-Mele model.} \textbf{a}. Schematic of the Rice-Mele model. \textbf{b}. A pumping cycle sketched (qualitatively) in $\delta$-$\Delta$ space.  
\textbf{c}. Schematic of the continuous Rice-Mele (cRM) pumping sequence. 
The pink shaded packet indicates the wave function of a particular atom initially localized at the unit cell $i$. 
The wave function shifts to right as the pumping proceeds and the atom moves to unit cell $i+1$ after one pumping cycle. The blue dashed curve and the green arrow indicate the harmonic confinement (not in scale) and an initial hole, respectively.  
}
\label{fig:RM_model}
\end{figure}

In this Letter, we report a realization of  Thouless' topological charge pump by exploiting the controllability of ultracold atoms in an optical superlattice. 
Differently from recent realizations of topological bands in two (spatial or synthetic) dimensions~\cite{Aidelsburger13, Miyake13, Jotzu14, Aidelsburger, Mancini:2015wl, Stuhl:2015wn},
our experiment explores the topology of a (1+1) dimensional adiabatic process,
in which a dynamically controllable one-dimensional optical superlattice is implemented following the proposal of Ref.~\cite{Wang}.
Topological pumping is seen as a shift of the  center of mass (CoM) of an atomic cloud measured with \emph{in situ} imaging. We  extract the Chern number of the pumping procedure from the average shift of the CoM per pumping cycle.
The topological nature of the pump is revealed by the clear dependence on the topology of the pumping trajectories in parameter space as to whether the trajectory is enclosing the degenerate point or not. 
Our work introduces a new experimental platform to study topological quantum phenomena in adiabatic driven systems.

In our experiments, an ultracold Fermi gas of ytterbium atoms $^{171}$Yb is prepared (see Methods A)
and loaded into a dynamically controlled optical superlattice.
Specifically, we construct a stationary lattice (short lattice) that has the period of 266~nm and a dynamical interferometric lattice (long lattice) that has the period of 532~nm whose phase is stabilized and controlled by a Michelson interferometer (see Methods B).
As a result, these laser beams create the required~\cite{Wang} time-dependent one-dimensional optical superlattice of the form
\begin{equation}
V(z,t) = -V_S(t)\cos^2\left(\frac{2\pi z}{d}\right)
-V_L(t)\cos^2\left(\frac{\pi z}{d}-\phi(t)\right),
\label{eq:lattice}
\end{equation}
where $d=532$~nm is the lattice constant of the superlattice, $V_S$ is the depth of the short lattice, $V_L$ the depth of the long lattice, and $\phi$ is the phase difference between the two lattices. In our experiments, $V_S$ and $V_L$ are controlled by the respective laser powers and $\phi$ by changing the optical path difference between the two interfering beams with a piezo-transducer (PZT)-mounted mirror, which enables us to sweep $\phi$ up to $\sim 11\pi$ corresponding to more than ten pumping cycles. In the following, we use the lattice constant $d$ as the unit of length and the recoil energy $E_R=h^2/(8md^2)$ as the unit of energy, 
where $h$ denotes Planck's constant and $m$ is the atomic mass of $^{174}$Yb (see Methods B).

\begin{figure}[tb]
\includegraphics[width=8.5cm,clip]{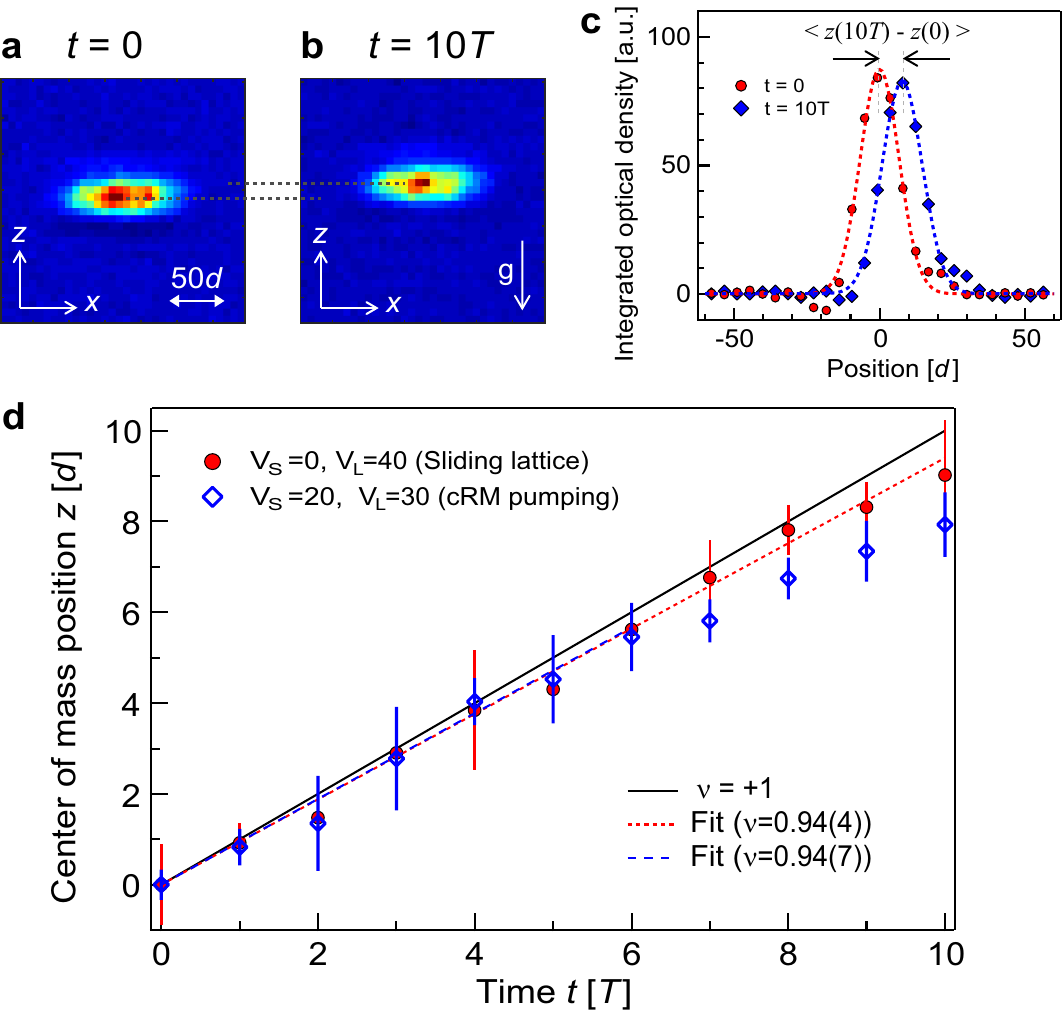}
\caption{
{\bf Observation of cRM pumping and sliding lattice pumping.}
\textbf{a-b}.  \textit{In situ} absorption images on the CCD before and after 10 cRM pumpings.
\textbf{c}. One-dimensional optical densities (integrated along the $x$ axis) 
before pumping (red circles, same data as \textbf{a})
and after 10 cRM pumping (blue diamonds, same data as \textbf{b}). \textbf{d}. The center of mass (CoM) of the atomic cloud after up to ten pumping cycles.  Red circles and blue open diamonds indicate the CoM shift of the sliding lattice and the cRM pumping lattice, respectively.  Error bars denote standard deviation of five independent measurements.
}
\label{fig:Pumping}
\end{figure}

We load $^{171}$Yb atoms into an array of one-dimensional optical superlattices, ensuring that they occupy the lowest energy band (see Supplementary Information S4),
and slowly sweep $\phi$ over time. The lattice potential returns to its initial configuration whenever $\phi$ changes by $\pi$, thus completing a pumping cycle. 
Since the lattice potential is periodic both in space and time, one can define energy bands,
the Bloch wavefunction $\ket{\psi_{k}(t)}=e^{ikz}\ket{u_{k}(t)}$, and corresponding topological invariants such as the Chern number $\nu$ in a $k$-$t$ Brillouin zone:
\begin{equation}
\nu = \frac{1}{2\pi} \int_{0}^{T}d t \int_{-\pi/d}^{\pi/d} d k\, \Omega(k,t), 
\label{eq:Chern}
\end{equation}
where $\Omega(k,t) = i(\braket{\partial_{t} u_{k}|\partial_{k} u_{k}}- \braket{\partial_{k} u_{k}|\partial_{t} u_{k}})$ is the Berry curvature (see Methods C) and $T$ the pumping period. 
We have ensured that the band gap never closes during the whole pumping procedure so ideally the atoms stay in the lowest band during the adiabatic pumping process. The phase sweep breaks time-reversal symmetry and the energy bands can acquire a non-zero Chern number $\nu$. 
The shift of the CoM of the atomic cloud in such a topologically non-trivial band after one pumping cycle is simply given by $\nu d$.

The ability to tune all parameters of the lattice potential (\ref{eq:lattice}) independently in a dynamic way offers the opportunity to realize various pumping protocols. In the absence of the static short lattice, $V(z,t)$ describes a simple sliding lattice which Thouless originally proposed ~\cite{Thouless}. Including the $V_S$ term, one realizes a double-well lattice illustrated in Fig.~\ref{fig:RM_model}. A pictorial understanding of this alternative pumping process is provided by the tight-binding Rice-Mele model~\cite{RM,Atala}, 
\begin{equation}\label{eq:RM}
\mathcal{\hat{H}}= \sum_i\left(-(J+\delta) \hat{a}_i^\dagger \hat{b}_i -(J-\delta)\hat{a}_i^\dagger \hat{b}_{i+1}+{\rm h.c.}
+\Delta (\hat{a}_i^\dagger \hat{a}_i-\hat{b}_i^\dagger \hat{b}_i) \right), 
\end{equation}
where $\hat{a}_i$ and $\hat{b}_i$ are  fermionic annihilation operators in the two sublattices of the $i$-th unit cell, $J\pm\delta$ is the tunnelling amplitude within and between unit cells, and $\Delta$ denotes a staggered on-site energy offset, as shown in Fig.~\ref{fig:RM_model}a.  We ignore the spin degree of freedom since we can neglect the interaction between the two spin components due to 
a very small $s$-wave scattering length~\cite{Kitagawa08}.

\begin{figure*}[htb]
\includegraphics[width=170mm]{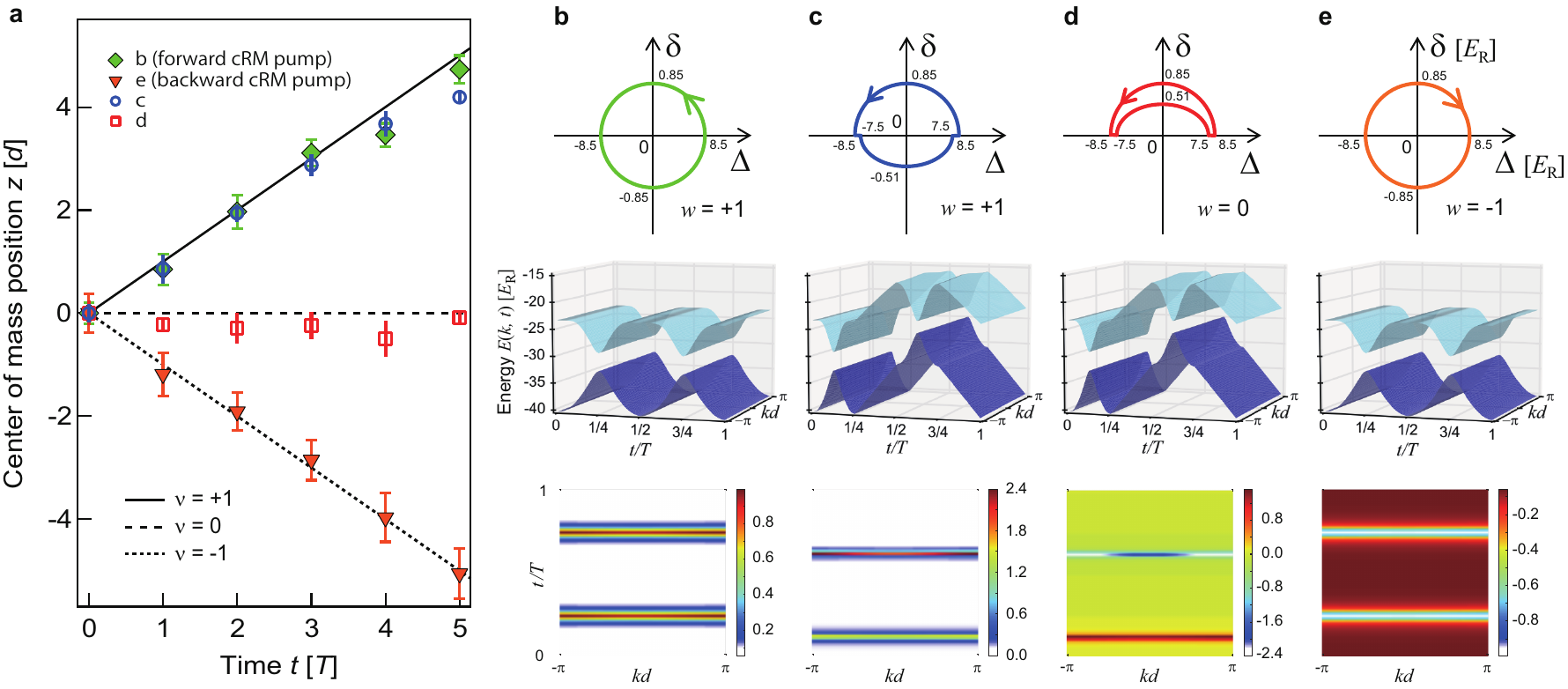}
\caption{
{\bf Topological aspects of cRM pumping}.
\textbf{a}. Charge pumped during a simple cRM pumping (\textbf{b}),
topologically-nontrivial pumping (\textbf{c}),
topologically-trivial pumping (\textbf{d}),
and negative sweep cRM pumping (\textbf{e}).
The vertical error bars denote the standard error of the mean of ten CoM measurements.
\textbf{b-e}. Pumping sequences in the $\delta$-$\Delta$ plane (top),
the corresponding band structures in the $k$-$t$ Brillouin zone (middle), 
and the Berry curvatures of the pumping cycles  (bottom). The indices $w$ in the top figures indicate the winding number of each trajectory around the origin.}
\label{fig:topology}
\end{figure*}

Figure~\ref{fig:RM_model}c shows the schematics of our ``continuous Rice-Mele'' (cRM) pumping sequence. 
Sweeping the phase linearly in time as $\phi(t)=\pi t/T$ the hopping amplitudes and on-site energies are modulated periodically.
Our~\emph{ab initio} calculation shows that the cRM pumping scheme used in the experiment is topologically equivalent to the Rice-Mele model for atoms that reside in the lowest energy band since the Chern numbers are the same~(see Supplementary Information S3). In the following, we will thus use the tight-binding Rice-Mele Hamiltonian to simplify the discussion of the pumping sequence
as a closed trajectory in the $\delta$-$\Delta$ parameter plane
(Fig.~\ref{fig:RM_model}b).
Note that, as shown in Fig.~\ref{fig:RM_model}c, our system has metallic edge states and thermal holes due to the combination of the trapping potential and finite temperature. We estimate the filling of the lattice is typically $\sim 0.7$ for each spin at the center of the trap. However, in the case of our deep optical lattice systems, the shift of the CoM of the atomic cloud still constitutes a quantized shift albeit these thermal and finite size effects (see Supplementary Information S2).

Figure~\ref{fig:Pumping} shows the main results of our pumping experiments. 
Our stable absorption imaging system with a charge-coupled-device (CCD) camera enables us to accurately measure the shift of the CoM of the atomic cloud after several pumping cycles (see Supplementary Information S5), as shown in Fig.~\ref{fig:Pumping}a and b. The period $T$ is fixed to 50~ms for the results shown in Fig.~\ref{fig:Pumping}.
One can clearly recognize the sizable CoM shift along the $z$-direction. We plot the {\it in situ} CoM positions of the atomic cloud after a few pumping cycles in Fig.~\ref{fig:Pumping}d.
The averaged CoM shift per cycle $\braket{z(t)-z(0)}/(td)$ of the cRM pumping with
$(V_{S}, V_{L})=(20, 30)E_{R}$ is evaluated to be $0.94(7)$ for $t\leq 6T$.
This provides a direct measurement of the Chern number of the occupied energy band,
which is consistent with the ideal value $\nu=1$. 
As a comparison, the observed average CoM shift per cycle of a sliding lattice $(V_S, V_L)=(0,40)E_{R}$ is $0.94(4)$, which is again close to the ideal value of $\nu=1$. Classically it is quite intuitive that the sliding lattice is able to transfer atoms because the potential minima are moving in space. However, even though the potential minima of the cRM pump $(V_{S}, V_{L})=(20, 30)E_{R}$ are not moving in space as shown in Fig.~\ref{fig:RM_model}c, the pumping is topologically equivalent because of the same Chern number of the occupied band. The cRM lattice has the same ability to transfer atoms residing in the lowest energy band, even though the pumping is achieved by a sequence of quantum tunneling events between the double-wells (see Supplementary Information S4). 
We attribute the saturating behavior of the cRM pumping for $t>6T$ to the effect of the harmonic confinement, whose variation can be comparable to the band gap for large CoM shift~\cite{PhysRevA.84.013608} (see Supplementary Information S6).

A striking feature of our pump is its \textit{topological} nature. In particular, the pumped amount in the Rice-Mele model~\cite{Xiao,Shen} is directly related to the topology of the trajectory in the $\delta$-$\Delta$ plane. It depends \emph{only} on the winding number $w$ of the trajectory that encloses the origin $\delta=\Delta=0$ (see Supplementary Information S3). Note that electron pumping in restricted nano-devices~\cite{Switkes,Blumenthal,Kaestner,Shilton} is not topological since there the amount of the  charge pumped per cycle instead depends on the area of the enclosed parameter space~\cite{Brouwer}, which is the geometry but not the topology of the trajectory. To highlight the topological nature of Rice-Mele pumping, we investigate four distinct pumping sequences with trajectories shown schematically in Figs.~\ref{fig:topology}b-e. In Fig.~\ref{fig:topology}a, we plot the CoM shifts of two cRM pumping schemes with $(V_S, V_L)=(20, 30)E_{R}$ (Fig.~\ref{fig:topology}b, e) and two amplitude-modified cRM pumping schemes (Fig.~\ref{fig:topology}c, d). Evidently, the sequence which does not wind around the origin (Fig.~\ref{fig:topology}d) results in no pumping, and those with winding trajectories (Fig.~\ref{fig:topology}b, c and e) result in finite pumping. 
Also the forward cRM pumping (Fig.~\ref{fig:topology}b) and the amplitude-modified cRM pumping (Fig.~\ref{fig:topology}c) exhibit almost same pumping behavior while the area enclosed by the trajectory of Fig.~\ref{fig:topology}c is actually smaller than that of Fig.~\ref{fig:topology}b.
This is direct evidence of the topological nature of the pump. Note that the band structure in the $k$-$t$ space of the non-trivial pumping sequence (Fig.~\ref{fig:topology}c) is \emph{identical} to that of the trivial pumping (Fig.~\ref{fig:topology}d). However, the Berry curvature and the Chern number of the lowest band are different.
This highlights the fact that the pumped charge is a topological quantity, which depends on the wave function but not on the band dispersions. Furthermore, we also performed the cRM pumping with a negative sweep of the phase $\phi(t)=-\pi t/T$, which corresponds to an opposite winding in $\delta$-$\Delta$ plane and the cloud is pumped to the opposite direction even though the band dispersion remains identical to that of the forward sweep pumping (Fig.~\ref{fig:topology}e). 

\begin{figure}[tb]
\includegraphics[width=8cm,clip]{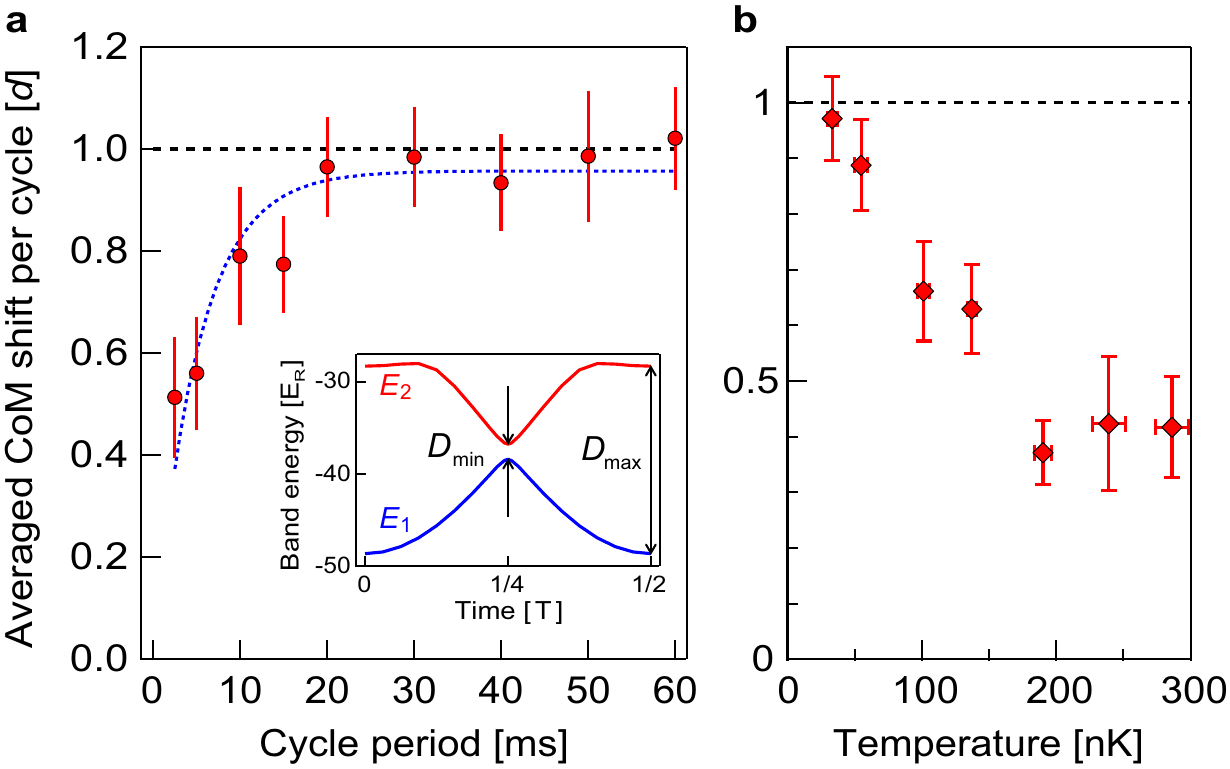}
\caption{
{\bf Conditions for quantum pumping.}
\textbf{a}. The averaged CoM shift per cycle (averaged after two cycles) versus the pumping period $T$.
The inset shows instantaneous band maxima $E_1$ of  the first band and minima $E_2$ of the second band
for $(V_S, V_L) = (30, 30)E_R$.
The pumped amount saturates as the pumping speed slows down.
The dotted curve shows an exponential fit with time constant $5.1(9)$~ms.
The error bars denote the 1$\sigma$ confidence bound derived from thirty CoM measurements.
\textbf{b}. Finite temperature effect in the cRM pumping.
The pumped amount (averaged after five cycles) approaches the ideal value as the temperature decreases. 
The vertical error bars denote the 1$\sigma$ confidence bound derived from ten CoM measurements. The horizontal axis is the initial temperature in the FORT before lattice loading,
evaluated from the Fermi-Dirac fitting in the degenerate regime
($T/T_F\lesssim 0.5$) and from the Gaussian fitting in the thermal regime.
The horizontal error bar indicates standard deviation of five independent temperature measurements.
}
\label{fig:SweepTemp}
\end{figure}

A crucial requirement of the topological Thouless pump is  adiabaticity,  requiring that the band gap never closes during the pumping process and that the atoms always remain in the lowest energy band. 
Figure \ref{fig:SweepTemp}a shows the pumping period dependence of the cRM pumping with the depths of $(V_S, V_L)=(30, 30)E_{R}$.
The data suggests that the averaged CoM shift per cycle reaches its ideal values if the pumping period $T$ is longer than $\sim 30$~ms, i.e., cycle period of $T>30$~ms is long enough to satisfy the adiabatic condition in this lattice potential.
This can be understood from considering Landau-Zener transitions to the higher band.
The instantaneous energy gap $D(t) = E_2-E_1$ changes in time as shown in the inset of Fig.~\ref{fig:SweepTemp}a. 
The diabatic transition probability $P$ is determined by the minimum band gap $D_{\rm min}$ and the maximum band gap $D_{\rm max}$ through the Landau-Zener formula $P=e^{-2\pi \Gamma}$,
where $\Gamma=(D_{\rm min}/2)^2/(\hbar \frac{d}{dt}D(t))$.
In our case, $D_{\rm min}=1.6 E_R$ and $D_{\rm max}=20 E_R$. 
Since the energy sweep speed $dD/dt$ is on the order of $8D_{\rm max}/T$, 
we find that $2\pi \Gamma \sim T/$(6~ms), which is consistent with the observed result.

We next check the temperature dependence of our pump. 
Figure~\ref{fig:SweepTemp}b shows the pumped amount as a function of the temperature of the gas before loading into the lattice of the depths of $(V_S, V_L)=(25,30)E_{R}$. 
 The temperature is tuned by changing the sympathetic evaporative cooling condition while keeping the same number of $^{171}$Yb atoms.
One can see that pumping reaches the ideal value at the lowest loading temperature in the FORT of 33(4)~nK, which corresponds to $0.24(4)T_F$, where $T_F$ is the Fermi temperature.
Assuming an adiabatic lattice loading, we estimate that the temperature
of atoms in the optical superlattice could be 65(15)~nK for this lowest loading temperature.
While the temperature given in Fig.~\ref{fig:SweepTemp}b is not the temperature in the lattice but that before lattice loading, the observed temperature dependence exhibits similar behavior to that discussed in Ref.~\cite{Wang}: pumping efficiency reaches its ideal value once the temperature becomes lower than the minimum band gap of $2.4 E_R$($=120$~nK) for this lattice depths.

Having demonstrated topological Thouless pumping using flexible optical superlattice setup, the scheme can be extended to even more novel setups. 
For example, by choosing a special lattice laser frequency, one can create a spin dependent superlattice~\cite{Mandel} which realizes $Z_2$-spin pumping~\cite{Fu}, a counterpart of $Z_2$ topological insulators. Another possibility is to change the ratio of the long lattice and short lattice wavelengths by tuning the angle of the interferometric lattice, which realizes a superlattice with incommensurate ratios and the Aubry-Andr\'e model~\cite{AA} with  fractional  pumping~\cite{Marra} and anomalous pumping~\cite{wei2015anomalous}. Furthermore, introducing interaction effect is feasible and will open a door for experimental exploration of the interplay of topological quantum phenomena and interaction and correlation effects.

\textit{Note added}. Recently, we became aware of similar work carried out by Lohse \textit{et al.}
~\cite{Lohse} observing the topological Thouless pumping with bosonic atoms in Mott insulator state.

\section{Methods}    
\subsection{Preparation of a degenerate Fermi gas of $^{171}$Yb}
Since $^{171}$Yb atoms have a very short $s$-wave scattering length of -0.15~nm~\cite{Kitagawa08},
we use sympathetic evaporative cooling with $^{173}$Yb atoms to obtain 
the degenerate Fermi gas of $^{171}$Yb~\cite{Taie10}.
After collecting Yb atoms in a magneto-optical trap using the intercombination transition (556~nm), 
the two isotopes are loaded into a crossed far-off-resonant trap (FORT) with 532~nm light.
Sympathetic evaporative cooling is performed by continuously decreasing the FORT trap depth.
After blowing away the $^{173}$Yb atoms by using a 556~nm laser which is resonant only for $^{173}$Yb,
we obtain a pure degenerate gas of $^{171}$Yb atoms
with two hyperfine spin components of $\ket{F=1/2, m_F=\pm 1/2}$.
The number of atoms for each spin is typically $9(1.5)\times 10^3$ for Fig.~\ref{fig:Pumping} and \ref{fig:SweepTemp}b, and $6(1.5)\times 10^3$ for Fig.~\ref{fig:topology} and \ref{fig:SweepTemp}a.
A typical temperature in the FORT before lattice loading is $T/T_F=0.26(8)$
at the end of the evaporation with the trap frequencies of the FORT of 
$(\omega_{x'}, \omega_{y'}, \omega_z)/2\pi=(170, 42, 153)$~Hz, 
where the $x'$- and $y'$- axes are tilted from the lattice axes ($x$ and $y$)
by 45$^{\circ}$.

\subsection{Setup for the optical superlattice}
Our one-dimensional optical superlattice setup is a part of our optical Lieb lattice system~\cite{Taie15}.
To stabilize the phase $\phi$ of the interfering 532~nm-spacing optical lattice (long lattice)
we construct a Michelson interferometer with frequency stabilized 507~nm laser,
whose optical paths are overlapped with the interfering lattice beams 
until separated from the lattice beams with dichroic mirrors before entering the experimental chamber.
The 507~nm laser beams are retro-reflected to form the interferometer just after the separation. 
We can control the phase $\phi$ by tuning the PZT of the retro-reflection mirror
of the $z$-axis 507~nm laser, 
keeping stabilization of the optical path lengths of the lattice and 507~nm beams 
via another PZT-mounted mirror in the common path as long as the phase sweep speed is smaller than $0.5 \pi$ rad/ms.
The short-term stability of the phase $\phi$ is estimated to be $0.007\pi$.
A phase drift of typically $0.05\pi$ per hour,
is not a problem since the pumping only depends on the phase difference before and after pumping.
The non-linearity of the PZT and the relative phase $\phi$ between the long lattice and the short lattice are calibrated via the matter wave interference pattern of a Bose-Einstein condensate (BEC) of $^{174}$Yb atoms released from the superlattice.
The depths of optical lattices are also calibrated via pulsed lattice with the BEC of $^{174}$Yb.

\subsection{Calculations of the band structure and Chern number}
To predict the pumped charge of the experimental pumping protocols, we calculate the band structures and the Chern numbers of the one-dimensional Hamiltonian $H(z, t) = -\frac{\hbar^{2}\nabla^{2}}{2m}+ V(z, t)$. The band structure is obtained by solving $H(z,t)\ket{ \psi_{k}(t) } = E(k, t) \ket{\psi_{k}(t) } $ in a plane wave basis. 
The Chern number is then calculated as 
\begin{equation}
\nu = \frac{1}{2\pi} \int_{0}^{T}d t \int_{0}^{2\pi/d} d k\, \Omega(k,t), 
\end{equation}
where $\Omega(k,t) = \partial_{t} A_{k} -\partial_{k}A_{t}$ is the Berry curvature and $A_{t(k)} =i \braket{ u_{k}(t) |\partial_{t(k)}|u_{k}(t)}$ is the Berry connection calculated using the periodic part of the Bloch wavefunction $\ket{ u_{k}(t) } = e^{-ikz} \ket{\psi_{k}(t)}$. 

\section*{Additional information}
Correspondence and requests for materials should be addressed to S. Nakajima.

\section*{Acknowledgements}
We thank N. Kawakami, S. Fujimoto, J. Ozaki, T. Fukui, I. Maruyama, Y. Hatsugai,  
and S. Nakamura for valuable discussions and A. Sawada for experimental assistance.
This work was supported by the Grant-in-Aid for Scientific Research of JSPS
(No.~25220711, No.~26247064, No.~24-1698),
and the Impulsing Paradigm Change through Disruptive Technologies (ImPACT) program. L.W. and M.T. were supported by ERC Advanced Grant SIMCOFE and by the Swiss National Science Foundation through the National Center of Competence in Research Quantum Science and Technology QSIT. L.W. and M.T. acknowledge Xi Dai for collaborations on the related topic. 

\section*{Author contributions}
S. N. and T. T. carried out experiments and the data analysis.
S. T. conceived the experimental techniques for the superlattice. T. I. and H. O. contributed to building up the superlattice setup.
L. W. carried out the theoretical calculation.
Y. T. conducted the whole experiment. All the authors contributed to the writing of the manuscript.

\section*{Competing financial interests}
The authors declare no competing financial interests.



\providecommand{\noopsort}[1]{}\providecommand{\singleletter}[1]{#1}%


\onecolumngrid
\clearpage
\begin{center}
\noindent\textbf{Supplementary Information for:}
\\\bigskip
\noindent\textbf{\large{Topological Thouless Pumping of Ultracold Fermions}}
\\\bigskip
Shuta~Nakajima$^{1}$, Takafumi~Tomita$^{1}$, Shintaro~Taie$^{1}$, Tomohiro~Ichinose$^{1}$, Hideki Ozawa$^{1}$, Lei~Wang$^{2}$, Matthias~Troyer$^{2}$, and Yoshiro Takahashi$^{1}$ \\
\vspace{0.1cm}
\small{$^1$ \emph{Department of Physics, Graduate School of Science, Kyoto University, Japan 606-8502}}\\
\small{$^2$ \emph{Theoretische Physik, ETH Zurich, 8093 Zurich, Switzerland}}
\end{center}
\bigskip
\bigskip
\twocolumngrid

\setcounter{figure}{0}
\renewcommand{\figurename}{FIG. S}
\newcommand{\figureref}[1]{S\ref{#1}}
\renewcommand{\bibnumfmt}[1]{[S#1]}
\renewcommand{\citenumfont}[1]{S#1}
\section*{S1. Polarization and Pumping in a 1D system}
Thouless~\cite{Thouless_S} calculated the current in a one-dimensional periodic potential using the Bloch wave function and derived the formula for the quantized particle transport. Here we summarize the derivation of the same formula as the shift of a Wannier functions~\cite{KS93_S, Xiao_S, Bernevig_S}, 
which is more intuitive and relevant for the experimental realization of a finite and trapped system. 

\begin{figure*}[tb]
 \includegraphics[width=12cm,clip]{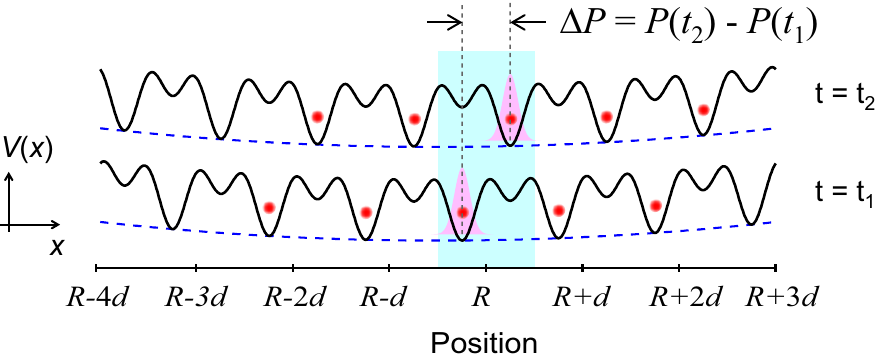}
\caption{
{\bf Thouless pumping as a change of the polarization.}
}
\label{fig:Polarization}
\end{figure*}

Figure~\figureref{fig:Polarization} shows the schematics of our pump.
We focus on a lattice site at $z=R$. Using the full Bloch function
of the lowest band $\ket{\psi_k}=e^{ikz}\ket{u_k}$, the Wannier function of the lowest band localized at the unit cell $R$ is given by
\begin{equation}
\ket{R}=\frac{1}{\sqrt{N}}\sum_{k=-\pi/d}^{\pi/d}e^{-ikR}\ket{\psi_k}
=\sqrt{\frac{d}{L}}\sum_{k=-\pi/d}^{\pi/d}e^{ik(z-R)}\ket{u_k}, \tag{S.1}
\end{equation}
where $N=L/d$ is the number of unit cells in the system,
$L$ the system length and $d$ the lattice constant. 
The expected shift of the Wannier center from the lattice site $R$, 
or ``polarization''~\cite{KS93_S} $P$
at time $t$ can be written as 
\begin{align}\label{eq:pol}
P(t)&=\braket{R(t)|z-R|R(t)}=\frac{d}{L}\sum_{k=-\pi/d}^{\pi/d}\bra{u_k(t)}i {\partial_{k}}\ket{u_k(t)} \nonumber \\
&=d\int_{-\pi/d}^{\pi/d}\frac{dk}{2\pi}A_k(k,t). \tag{S.2}
\end{align}
Here we used the relation  ${\partial_{k}}\ket{R}=0$.
The quantity $A_k(k, t)=i\bra{u_k(t)}{\partial_{k}}\ket{u_k(t)}$ is called the Berry connection and the integration on the right hand side of Eq.~(\ref{eq:pol}) is known as the Zak phase~\cite{Zak_S}.
Thus, the spatial shift of the Wannier function from $t=t_1$ to $t=t_2$
or the change of the polarization is given by  
\begin{equation}
\Delta P = P(t_2)-P(t_1)=d\int_{-\pi/d}^{\pi/d}\frac{dk}{2\pi}\left[ A_k(k,t_2)-A_k(k,t_1) \right ]. \tag{S.3}
\end{equation}
Using Stokes's formula, this can be written as
\begin{equation}\label{eq:polarization}
\Delta P = -d\int_{-\pi/d}^{\pi/d}\frac{dk}{2\pi} \int_{t_1}^{t_2} dt 
\left({\partial_{k}} A_t(k,t)-{\partial_{t}} A_k(k,t)\right), \tag{S.4}
\end{equation}
where $A_t(k, t)=i\bra{u_k(t)}{\partial_{t}}\ket{u_k(t)}$.
Then the shift of the Wannier function after one pumping cycle is
\begin{equation}
\Delta P = d \frac{i}{2\pi} \int_0^T \hspace{-2pt} dt \int_{-\pi/d}^{\pi/d} \hspace{-2pt} dk
\left[
\Braket{\frac{\partial u_{k}}{\partial t} | \frac{\partial u_{k}}{\partial k}} - \Braket{\frac{\partial u_k}{\partial k} | \frac{\partial u_k}{\partial t}} 
\right], \tag{S.5}
\end{equation}
where $\ket{\frac{\partial u_k}{\partial k}}={\partial_{k}}\ket{u_k}$.
This formula is essentially the same as the one derived by Thouless \cite{Thouless_S}.
Such an integral over two parameters in which the Hamiltonian is periodic should give an integer multiple of $2\pi i$, as discussed in TKNN's paper~\cite{TKNN_S}. 
Therefore, the shift of the cloud after one pumping cycle is simply given as $\Delta P = \nu d$, 
where $\nu$ is the first Chern number. In other words, the shift per cycle divided by $d$ gives the Chern number.

In the above derivation, we assume that the localized function can be written in the Wannier function consisting of the lowest band Bloch function.
In the interacting case, a  filled band is required for well-defined Wannier functions.
However, in our experiments, we can ignore interactions and
even if we have thermal holes at the beginning of the pumping,
each Wannier functions move independently and the center of mass of the cloud is robustly pumped as we observed (see also discussions in S2).    


\begin{figure*}[tb]
\includegraphics[width=15cm,clip]{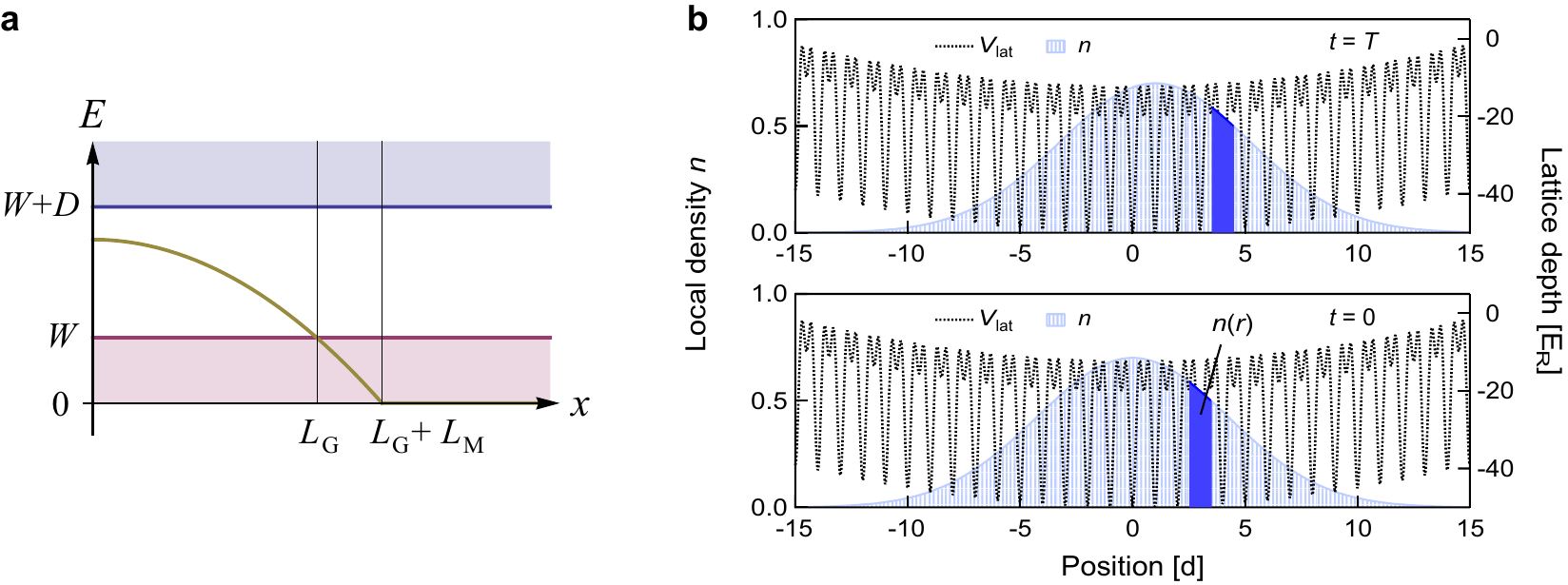}
\caption{{\bf Effect of harmonic confinement under the LDA.}
\textbf{a}. 
LDA estimation for the length of topologically protected (gapped) region $L_G$
and metallic wings $L_M$. $W$ and $D$ denote the band width of
the lowest band and the band gap to the higher band respectively.
Golden line indicates the local chemical potential in the trap.
\textbf{b}. Thouless pumping with a harmonic confinement.  
The dotted curve indicates the initial loading lattice $(V_S, V_L)=(20, 30)E_R$ with a harmonic confinement of $\omega_T=2\pi \times 140$~Hz.
Because the amount of the pumped charge is proportional to the local density $n(r)$ (blue solid), the whole density distribution (blue shaded) after one pumping cycle ($t=T$, upper panel) is almost the same as initial one ($t=0$, lower panel) except its CoM shift of $d$.}
\label{fig:metallic}
\end{figure*}

\section*{S2. Effect of the metallic edge state and thermal holes}
In this section we discuss the effect of the metallic edge state and thermal holes, 
which are inevitable for the atoms at a finite temperature in a harmonic confinement.

Since the trapping potential varies much slower than the optical lattice potential, it provide a spatially varying local chemical potential $\mu(x)=\mu_0-\frac{1}{2}m\omega_T^2 x^2$ under the local-density-approximation (LDA), where $\omega_T$ is the trap frequency of the harmonic confinement.
Depending on the relative position of the local chemical potential and the energy band of the lattice, the system is either gapful or gapless. In below we estimate the length of topologically protected (gapped) region ($L_G$)
and metallic wings ($L_M$). We start our discussion from almost filled condition, 
namely, the atoms are filled up to just below the second band (see Fig.~S\ref{fig:metallic}a).
In this case, the chemical potential in the center of the trap reaches $W+D$,
where $W$ and $D$ as the band width of the lowest band and the band gap
to the higher band, respectively. Then one obtains
\begin{align}
W+D&=\mu(0)=\mu_0 \notag \\ 
W&=\mu(L_G)=\mu_0-\frac{1}{2}m\omega_T^2 L_G^2 \tag{S.6} \\
0&=\mu(L_G+L_M)=\mu_0-\frac{1}{2}m\omega_T^2 (L_G+L_M)^2. \notag
\end{align}
One has $L_M/L_G=\sqrt{W+D}/\sqrt{D}-1$. The ratio is proportional to $W/D$ when $W/D \ll 1$. Notice that this ratio does not depend on the strength of the trap $\omega_T$ although $L_G$ itself depends on $\omega_T$.

Now we discuss our cRM pumping lattice with the depths of $(V_S, V_L)=(20, 30)E_R$.
In the atom loading condition ($\phi=0$), the band gap $D=17 E_R$ and the band width $W=0.00020 E_R$. Thus we obtain $L_M/L_G=5.9 \times 10^{-6}$, which is completely negligible.
Even in the minimum gap condition ($\phi=\pi/4$),    
the band width $W=0.0125 E_R$ is much smaller than the band gap $D=3.4 E_R$ 
and thus we obtain $L_M/L_G=1.8 \times 10^{-3}$, 
which is again negligible (see also Fig.~S\ref{fig:BandCalc}a).
In the case of a finite temperature or lower filling, 
the center chemical potential $\mu(0)$ becomes smaller than $W+D$.
However, our numerical calculation shows that $\mu(0)=1.1E_R$ in our typical temperature in the FORT of $T/T_F=0.26$,
which is still much larger than the band width $W$ and we can neglect the contribution of the metallic 
edge state.

\begin{figure*}[tb]
 \includegraphics[width=16cm,clip]{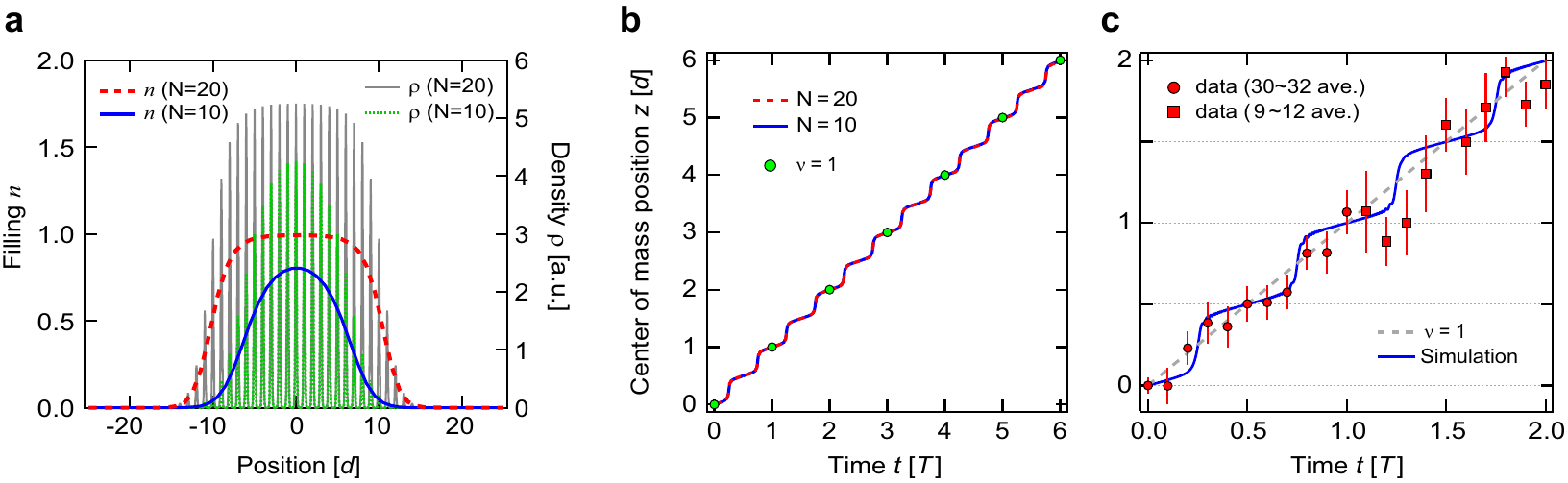}
\caption{{\bf Numerical simulation with harmonic confinement and thermal holes.}
\textbf{a}. One dimensional density distribution and filling of the cRM lattice for the case of $N=10$ and $N=20$ with a finite temperature of $1.2E_R$ and a harmonic confinement of $\hbar \omega_T=0.15E_R$.
\textbf{b}. Dynamical simulation of the cRM pumping with depths of $(V_S, V_L)=(20, 26)E_R$
with the finite temperature and the harmonic confinement. The blue solid line and the red dashed line are the simulation result of $N=10$ and $N=20$, respectively.
The green circles indicate the ideal pumping with $\nu=1$.
\textbf{c}. The CoM of the atomic cloud in the cRM pumping with depths of $(V_S, V_L)=(20, 26)E_R$.
The error bars of the data in the region of $0\leq t \leq T$ (red circles) represent 
the standard error of the mean of 30 to 32 independent measurements
and those in region of $T< t \leq 2T$ (red squares) the standard error of the mean of 9 to 12 independent measurements.
}
\label{fig:simulation}
\end{figure*}

Here we note that we have considerable thermal holes in our band due to a finite temperature effect, namely, the filling is smaller than unity (one particle per unit cell) for each spin. We estimate the filling of the lattice $n\sim 0.7$ for each spin at the center of the trap from the initial temperature in the FORT ($T/T_F=0.26$) assuming adiabatic lattice loading with $N=5\times 10^{3}$ atoms for each spin.
Note that this lower filling than unity or the existence of thermal holes does not come from the thermal excitation to the higher band. Actually, our numerical calculation suggests that more than 99\% of atoms are loaded into the lowest band since the evaluated temperature in the lattice ($\sim 1.2E_R$) is much smaller than the band gap of the initial loading lattice.
In fact, we observed that most of the atoms are mapped to the 1st Brillouin zone in our band mapping measurements, which corresponds to about 90\% population in the lowest band.
Below we show that, even under this situation, the shift of the CoM of the atomic cloud provides a Chern number $\nu$ given in Eq.(2) of a main text.
Going back to Thouless's original discussion, the pumped current after one pumping cycle $C$ is given by
\begin{equation}
C = \frac{d}{2\pi} \int_0^T dt \int_{-\pi/d}^{\pi/d} dk f(\epsilon_k) \Omega(k,t), \tag{S.7}
\end{equation}
where $f(\epsilon)=1/(e^{(\epsilon-\mu)/k_BT}+1)$ is the Fermi-Dirac distribution function,
$\epsilon_k$ the band dispersion and $\Omega(k,t)$ the Berry curvature. 
In general case, therefore, we have to integrate $\Omega(k,t)$ with a weight of $f(\epsilon_k)$.
If the band is totally filled, $f(\epsilon)=1$ and $C=\nu d$, where $\nu$ is the Chern number of Eq.(2). 
In the current situation, we can also treat $f(\epsilon_k)$ as almost uniform in the $k$-space due to the following reason, and can relate $C$ with the Chern number $\nu$.
Let's consider the distribution function at a certain point $x=r$.
The distribution function under the LDA is given by
\begin{equation}
f(\epsilon_k, r)=\frac{1}{e^{(\epsilon_k-\mu(r))/k_BT}+1}. \tag{S.8}
\end{equation}
Here $\epsilon_k$ spans from zero at $k=0$ to $W$ at $k=\pm \pi/d$
(we set the origin of the energy as the lowest band energy (the same origin of our chemical potential calculation)).
Except for the small metallic region, the band width $W$ is much smaller than $\mu(r)$ and temperature $k_BT$, as discussed above, and thus we can simply treat $f(\epsilon_k,r)=f(0, r)=const.$ for any $k$
(This situation is similar to the case of Ref.~\cite{Aidelsburger_S}, where thermal boson is loaded into lattice to achieve uniform population of energy bands). 
Therefore, the pumped charge at $x=r$ is given by
\begin{align}\label{eq:local}
C(r) &= f(0,r) \frac{d}{2\pi} \int_0^T dt \int_{-\pi/d}^{\pi/d} dk \Omega(k,t) = f(0, r)\nu d \nonumber \\
&=n(r) \nu d, \tag{S.9}
\end{align}
where $f(0,r)=n(r)$ is the local filling at the position $r$.
Note that this ``local'' pumped charge after one pumping cycle is also given by the integration of the local current $j(r,t)$ as $C(r)=\int_0^T j(r,t)dt$. By integrating the equation of continuity $\frac{\partial n(x,t)}{\partial t}+\frac{\partial j(x,t)}{\partial x}=0$ from $t=0$ to $t=T$,
one obtains $n(x,T)=n(x,0)-\frac{\partial n}{\partial x}\nu d$.
Therefore, the CoM shift after one pumping cycle is given by
\begin{align}
\Delta x_{\rm CoM}&=\frac{1}{N}\int_{-\infty}^{\infty}xn(x,T)dx-\frac{1}{N}\int_{-\infty}^{\infty}xn(x,0)dx \nonumber \\
&=-\frac{\nu d}{N}\int_{-\infty}^{\infty}x\frac{\partial n(x,0)}{\partial x}dx=\nu d, \tag{S.10}
\end{align}
where $N=\int_{-\infty}^{\infty}n(x,0)dx$ is the total atom number in a tube.
Thus, the shift of the CoM of the density distribution is simply given by $\nu d$
as we observed (see Fig.~S\ref{fig:metallic}b).
Note that this argument fails in the region in which $W$ becomes comparable to the local chemical potential $\mu(r)$, namely, the metallic wing. 
However, as we discussed above, contribution of such region is quite small in our deep lattice setup.
Due to these reasons, we can observe robust topological pumping 
even with the harmonic confinement and thermal holes.

\begin{figure*}[tb]
 \includegraphics[width=12cm,clip]{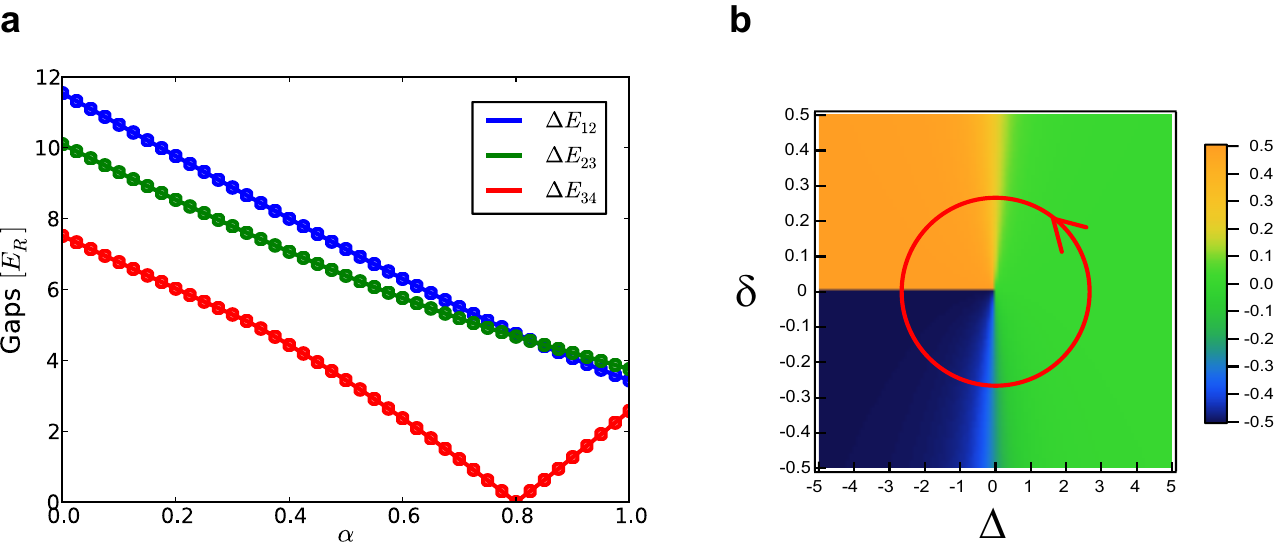}
\caption{
{\bf Band gaps and Zak phases.} 
\textbf{a}. Band gaps of the interpolated pumping between $(V_{S}, V_{L})=(0,40)E_{R}$ and $(20, 30)E_{R}$.
$\Delta E_{ij}$ denotes the band gap between energy bands $E_i$ and $E_j$. 
\textbf{b}. Zak phase divided by $2\pi$ in the plane of $\Delta$ and $\delta$ in the Rice-Mele model.
The line of discontinuity can be chosen anywhere depending on the gauge phase choice of the eigenstate.
The Zak phase or the polarization changes significantly when $\Delta$ changes the sign (See also discussion in S4). 
}
\label{fig:TT}
\end{figure*}

To quantitatively check the validity of the above discussion, we also perform \textit{ab initio} simulations of the pumping dynamics. 
Figure~S\ref{fig:simulation} shows a numerical simulation of the one dimensional cRM pumping with the lattice potentials of $(V_S, V_L)=(20, 26)E_R$ containing $N=20$ or $N=10$ spinless fermions in a tube
calculated by the same methods in Ref.~\cite{Wang_S}. In this calculation, we set the temperature of the atoms in the lattice $1.2E_R\sim 60$~nK, the trapping frequency $0.15E_R \sim 150$~Hz, and the cycle $50\hbar/E_R\sim 50$~ms.
In the case of $N=10$, the initial density doesn't reach unity
even at the center of the cloud as shown in Fig.~S\ref{fig:simulation}a. 
However, our numerical simulation suggests that both $N=10$ and $N=20$ show the same CoM dynamics (Fig.~S\ref{fig:simulation}b).
Figure~S\ref{fig:simulation}c compares the numerical simulation with the CoM shift measured in experiments (averaged over all tubes), which shows nice agreement especially in the initial stage of the pumping.


\section*{S3. Topological aspects of the continuous Rice-Mele model}
The simple sliding potential has Chern number $\nu= 1$ for each energy band, therefore the pumped amount is proportional to the band occupation. This is a manifestation of the Galilean invariance of the lattice~\cite{Kitagawa:2010bu}. By linearly tuning the lattice depth from $(V_{S},V_{L}) = (0, 40)E_{R}$ 
to the cRM pumping $(V_{S},V_{L}) = (20, 30)E_{R}$ with an interpolation parameter $\alpha\in[0, 1]$, the lowest two bands are adiabatically connected without gap closing as shown in Fig.~S\ref{fig:TT}a. The lowest two energy bands of the cRM pumping therefore also have Chern number $\nu=1$. However, a gap closes  between the higher bands, and the third band of  $(V_{S},V_{L}) = (20, 30)E_{R}$ lattice has Chern number $\nu = -1$. The above discussion also shows that the topological equivalence of the cRM pumping adopted in the experiment and the two-band tight-binding Rice-Mele model~\cite{RM_S} only holds for the lowest band, since the second energy band of the tight binding Rice-Mele model has Chern number $\nu = -1$.

For the Rice-Mele model, the pumping cycle defines a mapping from time to a closed path on the $\Delta-\delta$ plane.
Focussing on the lowest energy band, we now show that the pumped amount is related to the winding number of the path in the $\Delta-\delta$ plane~\cite{Xiao_S}. In the periodic gauge the first term in Eq.~(\ref{eq:polarization}) vanishes, therefore the change of the polarization is given by the path integral over the Zak phase, which is a smooth function except on the branch cut with a discontinuity of the size one. Figure~S\ref{fig:TT}b clearly shows that the resulting change of the polarization will be finite only for paths that wind around $\Delta=\delta=0$. 

\begin{figure*}[tb]
 \includegraphics[width=12cm,clip]{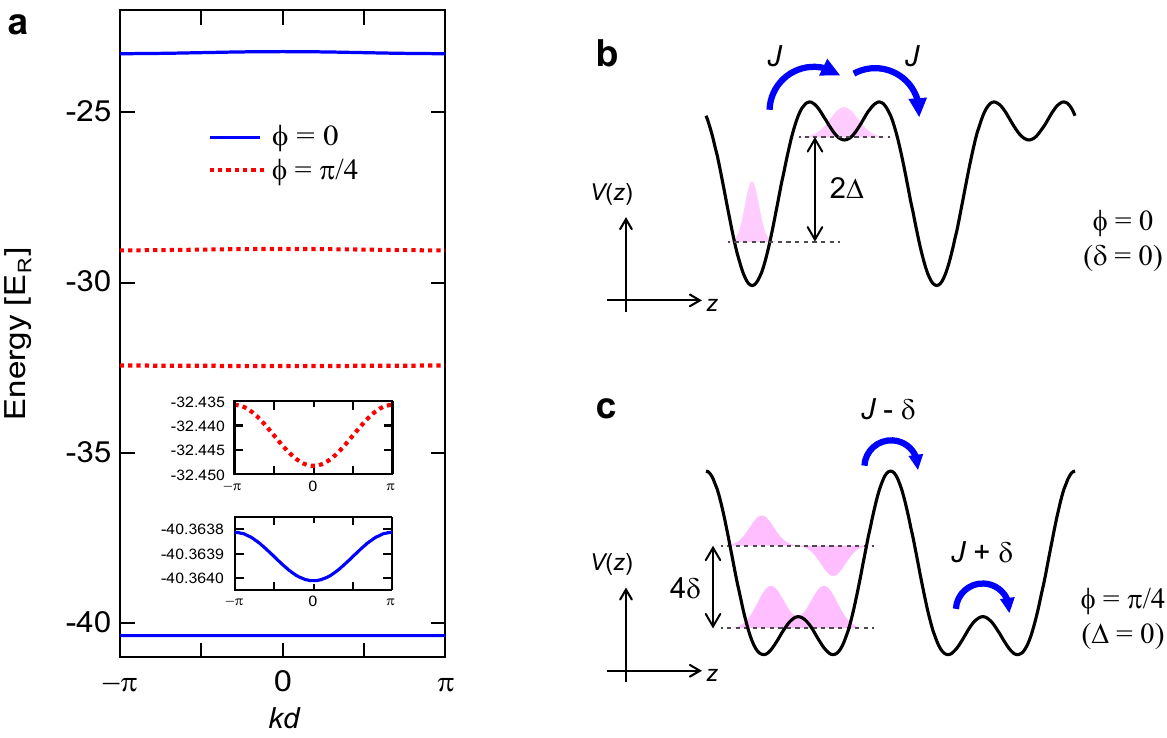}
\caption{
{\bf Estimate the Rice-Mele parameters from the band calculation.}
\textbf{a}. Calculated energy dispersions of the first and second band of the cRM pumping lattice with $(V_S, V_L)=(20, 30)E_R$. The blue solid and the red dashed line are the band structures at $\phi=0$ and $\phi=\pi/4$, respectively. The insets are the enlarged drawings of the each first band.
\textbf{b}. and \textbf{c}. Schematics of the cRM pumping lattice at $\phi=0$ and $\phi=\pi/4$,
respectively. The pink shaded figures schematically represent the Wannier functions of the first and second band.
}
\label{fig:BandCalc}
\end{figure*}
\begin{figure*}[tb]
 \includegraphics[width=16cm,clip]{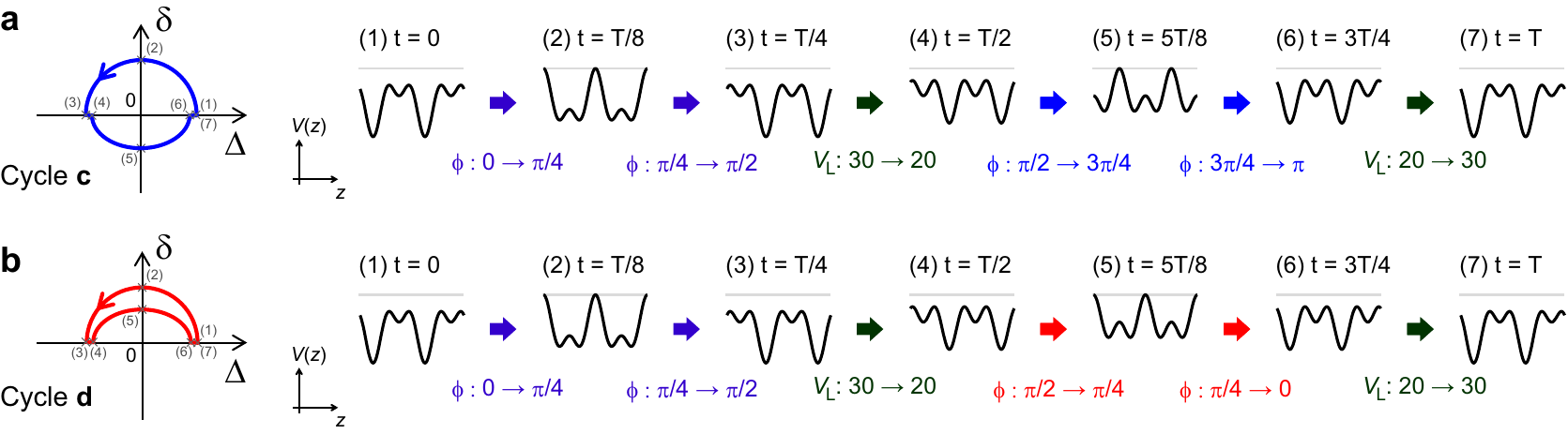}
\caption{
{\bf Schematic pumping sequences of the cycles of \textbf{c} and \textbf{d} in Fig.~3 in the main text.} 
The phase $\phi$ and the depth $V_L$ of the long lattice are linearly swept alternatingly. The short lattice depth $V_S$ is fixed to $30~E_R$ and the pumping period $T$ is 100~ms for both cycles. 
}
\label{fig:cycleCD}
\end{figure*}

In the experimental setup, especially if $V_L$ is comparable or larger than $V_S$, the modulated tunneling amplitude $J\pm \delta$ changes drastically in the pumping sequence
and $J$ is not constant as in the Rice-Mele model. Moreover, hoppings beyond nearest-neighbors may be required to fully capture the band structures. However, these effects do not close the gap and thus do not change the topology of pumping sequence -- a manifestation of the topological robustness of this system.
Therefore, we can still characterize the pumping using $\delta$ and $\Delta$ of the Rice-Mele model even in the experimental situations. 
To determine the Rice-Mele parameter ($\delta, \Delta$) in Fig.~3b-e in the main text, we use 
the calculated band structures of our cRM pumping lattice and the analytical form of the dispersion relation of the tight-binding Rice-Mele Hamiltonian of Eq.(3) in the main text:
\begin{equation}
\epsilon_{\pm}(k)=\pm\sqrt{4J^2\cos^2(kd/2)+4\delta^2\sin^2(kd/2)+\Delta^2}. \tag{S.11}
\end{equation}
As shown in Fig.~S\ref{fig:BandCalc}a, the actual band dispersion doesn't have the particle-hole symmetry $\epsilon_-(k)=-\epsilon_+(k)$. Therefore the tight-binding Rice-Mele Hamiltonian of Eq.(3) is not enough to characterize both the lowest and the second band as mentioned above.
However, characterising the lowest band structure of our cRM pumping lattice with the tight-binding Rice-Mele parameters $\delta$ and $\Delta$ is still reasonable as follows.
In the case of $\phi=0$, namely the staggered lattice (Fig.~S\ref{fig:BandCalc}b), 
the modulation of the tunneling $\delta=0$ and the minimum band gap at $k=\pi/d$ gives $\epsilon_+(\pi/d)-\epsilon_-(\pi/d)=2\Delta$.
As one can see in Fig.~S\ref{fig:BandCalc}a, the band width of the 1st band 
$W=\epsilon_-(\pi/d)-\epsilon_-(0)=\sqrt{4J^2+\Delta^2}-\Delta$
is much smaller than the gap, i.e., we can approximate $W=2J^2/\Delta$.
This result is reasonable since the band width of the lowest band gives the inter unit cell tunneling, 
which is given by the second order (the next-nearest) tunneling $\sim J^2/2\Delta$ in this situation (Fig.~S\ref{fig:BandCalc}b). 
In the case of $\phi=\pi/4$, namely the double-well lattice (Fig.~S\ref{fig:BandCalc}c),
the staggered energy offset $\Delta=0$ and the minimum band gap at $k=\pi/d$ gives
$\epsilon_+(\pi/d)-\epsilon_-(\pi/d)=4\delta$.
The band width of the 1st band
$W=\epsilon_-(\pi/d)-\epsilon_-(0)=4(J-\delta)$
again gives inter unit cell tunneling $J-\delta$ (Fig.~S\ref{fig:BandCalc}c).
The tunneling inside of the double well is described by $J+\delta \approx 2\delta$, which is determined by the band gap.
Thus, in the lowest band, the $\delta$ and $\Delta$ still have the same physical meaning of the Rice-Mele model.
For example, we can evaluate the tunneling $J$ and the Rice-Mele parameters $(\delta, \Delta)$ of our cRM pumping with the depths of $(20, 30)E_R$ as
$(J, \delta, \Delta)=(0.0290, 0, 8.54)E_R$ for $(\phi=0)$ and $(0.851, 0.845, 0)E_R$ for $(\phi=\pi/4)$.

\begin{figure*}[tb]
 \includegraphics[width=15cm,clip]{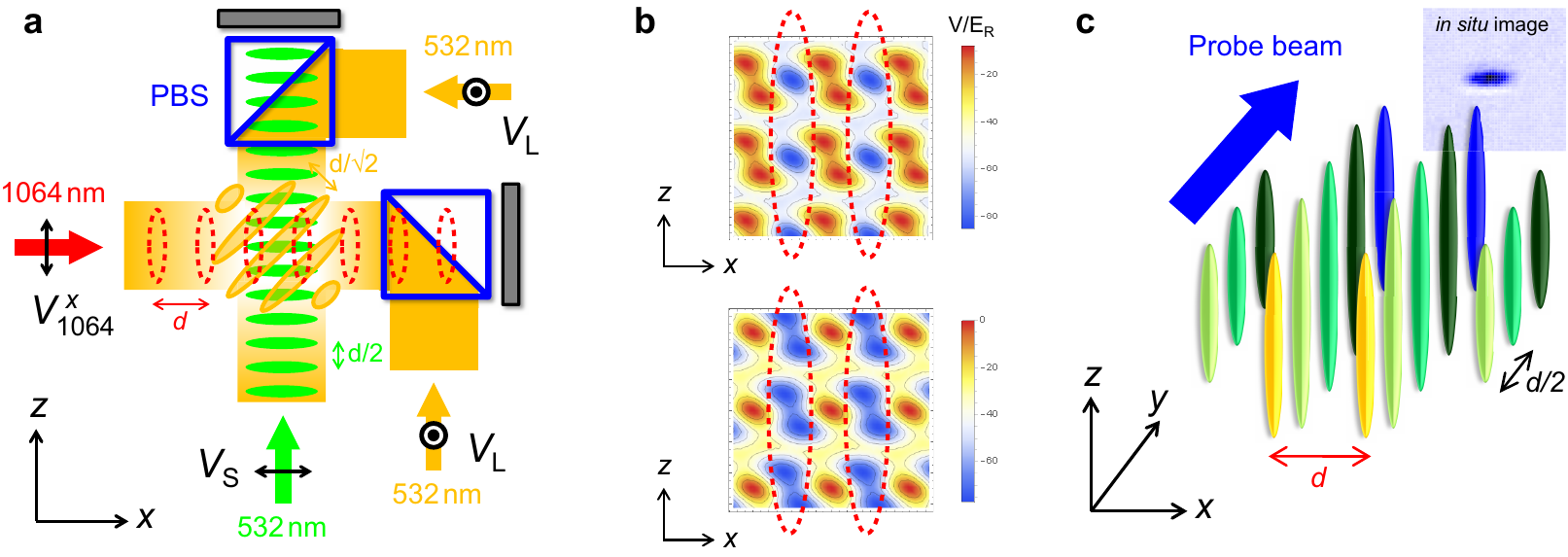}
\caption{
{\bf Configuration of our superlattice setup.}
\textbf{a}. Laser configuration for the experimental realization of
a one-dimensional optical superlattice. Black arrows indicate polarizations of the
lattice beams.
\textbf{b}. Lattice potential for $(V_S, V_L, V_{1064}^x)=(20, 30, 35)$ at $\phi=0$ (upper) and $\pi/2$ (lower). The red dashed ovals indicate one-dimensional superlattice regions. 
\textbf{c}. Schematic view of our superlattice tubes array and the imaging axis.
}
\label{fig:LatticeLasers}
\end{figure*}
\begin{figure*}[tb]
 \includegraphics[width=15cm,clip]{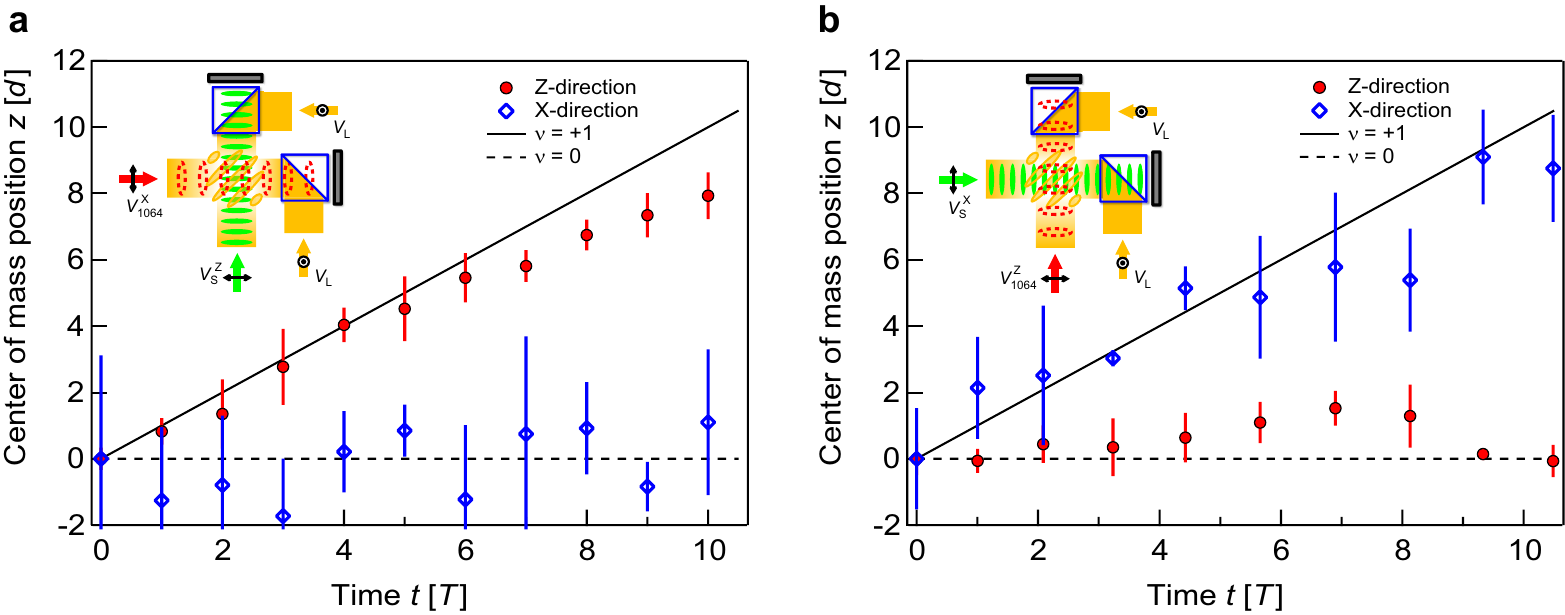}
\caption{
{\bf Absence of the shift of the cloud along the perpendicular direction.}
\textbf{a}. The CoM shift of the cRM pumping along the pumping direction ($z$, red circles)
and the perpendicular direction ($x$, blue open diamonds) in the $z$-axis superlattice configuration (inset of \textbf{a}). 
\textbf{b}. The CoM shift of the cRM pumping along the pumping direction ($x$, blue open diamonds)
and the perpendicular direction ($z$, red circles) in the $x$-axis superlattice configuration (inset of \textbf{b}).
The lattice depths are $(V_S, V_L)=(20, 30)E_R$ for both \textbf{a} and \textbf{b}.
}
\label{fig:reverse}
\end{figure*}

The Rice-Mele parameters on the trajectories of Fig.~3b-e in the main text are also evaluated like above. Note that the pumping sequence of the 
amplitude-modified cRM pumping (Fig.~3c, d) are the combinations of the cRM pumping of $(V_S,V_L)=(20, 30)E_R$ and $(20, 20)E_R$ (Fig.~S\ref{fig:cycleCD}).
The difference between these two sequences is the direction of the phase sweep. In the topologically-non-trivial cycle c (Fig.~S\ref{fig:cycleCD}a), the direction of the phase sweep doesn't change while its lattice depth $V_L$ are modified during one pumping sequence. 
On the other hand, the direction of the phase sweep of the topologically-trivial cycle d (Fig.~S\ref{fig:cycleCD}b) is reversed at $t=T/2$.
If we don't modify the amplitude $V_L$ in the sequence d,
this will simply be a ``back and forth'' moving lattice, which gives no pumping obviously.


\section*{S4. Details of dynamically-controlled superlattice setup}
In this section we explain the details of our dynamically-controlled optical superlattice setup and discuss the band filling.
The lattice laser configuration in the $z$-$x$ plane is shown in Fig.~S\ref{fig:LatticeLasers}a. 
The short lattice (a 266~nm-spacing optical lattice) in the $z$-direction is created by retro-reflection of a 532~nm laser (green arrow).
The long lattice (a 532~nm-spacing optical lattice with a time-dependent phase)
is created by the interference of two 532~nm laser beams (yellow arrows), stabilized by a Michelson interferometer. 
As shown in Fig.~S\ref{fig:LatticeLasers}a, 
the direction of the wavevector of this interferometric lattice or a ``diagonal lattice'' is not
$z$-axis but ($\hat{\bm{z}}-\hat{\bm{x}}$) axis
and the lattice has the lattice constant of $d/\sqrt{2}=376$~nm.
Moreover, the phase of the diagonal lattice moves to ($\hat{\bm{z}}-\hat{\bm{x}}$) direction
as we sweep the PZT in the Michelson interferometer. 
However, thanks to the $x$-direction confinement lattice $V_{1064}^x$ created by a standing wave of a 1064~nm laser beam indicated by the red arrow in
Fig.~S\ref{fig:LatticeLasers}a,
the hopping for the $x$-direction is suppressed and
we can regard the $z$-direction projected component of this moving diagonal lattice
as a moving long lattice along the $z$-direction, which creates a one-dimensional superlattice potential for the $z$-direction (Fig.~S\ref{fig:LatticeLasers}b).

\begin{figure*}[tb]
 \includegraphics[width=15cm,clip]{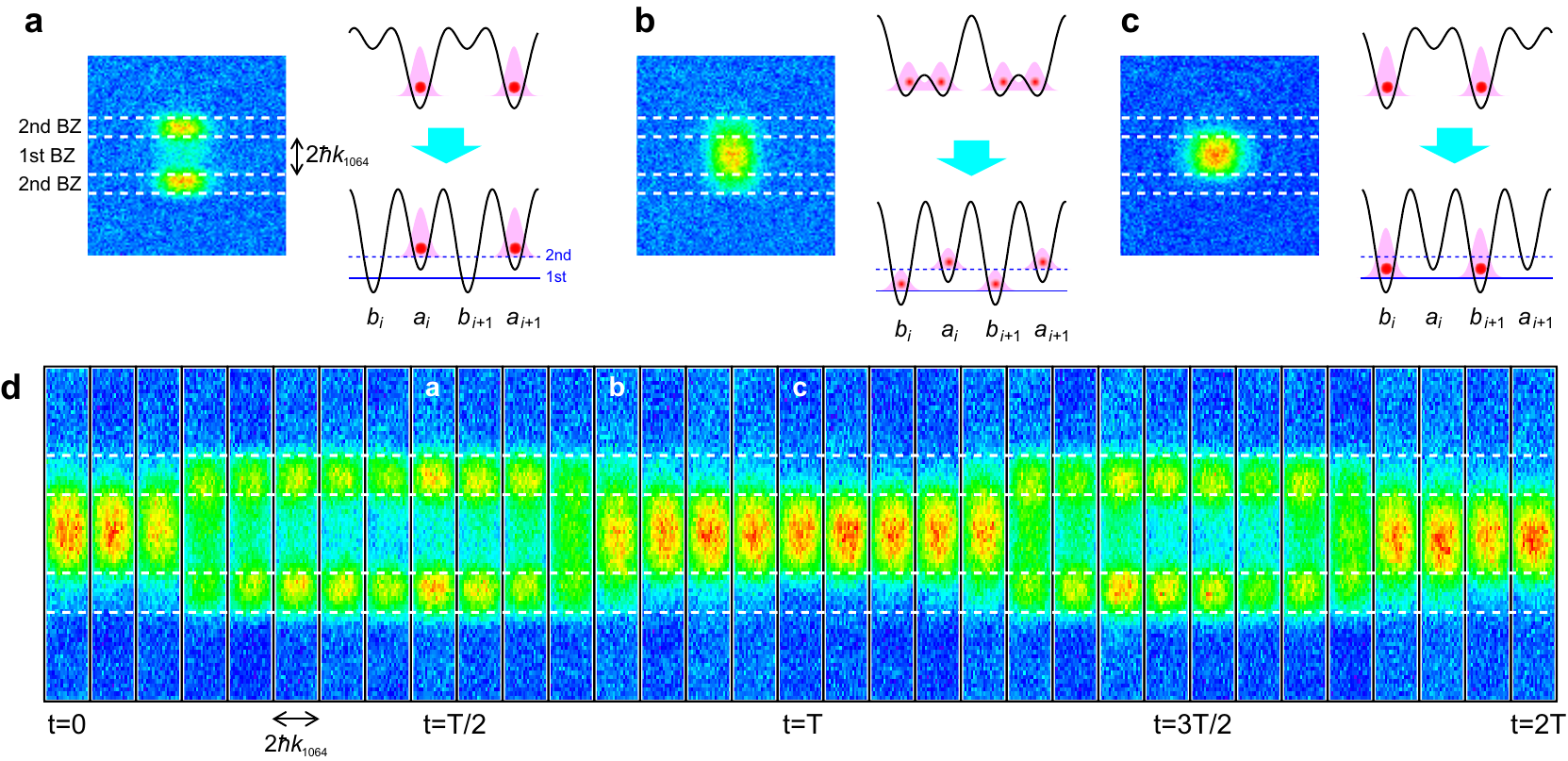}
\caption{
{\bf Confirming occupancy of the lowest band.}
\textbf{a}. Sublattice mapping technique. If a wave function is localized at sublattice site $a_i$,
the wave function is projected to the 2nd band of the mapping lattice and thus the occupation of the sublattice site $a_i$ can be 
measured via band mapping technique as the occupation of the 2nd band.
The left image in \textbf{a} is a band mapped absorption image taken after 10 ms time-of-flight (TOF)
(five times averaged).
\textbf{b}. If the wave function is delocalized over both $a_i$ and $b_i$, the wave function is mapped to both 1st and 2nd bands.
\textbf{c}. If a wave function is localized at sublattice site $b_i$, its occupation is mapped to the 1st Brillouin zone.
\textbf{d}. Sublattice mapping measurements (five times averaged). 
The horizontal scale of the TOF images are half of the TOF images in \textbf{a}-\textbf{c}.  
}
\label{fig:BM}
\end{figure*}

The setup contains another 532~nm laser to suppress the hopping for the $y$-direction ($V_{\bot}^y$)
and realizes an array of one-dimensional superlattice tubes (Fig.~S\ref{fig:LatticeLasers}c).
The depth of these vertical confinement lattices are 
$V_{1064}^x = 35E_R$ and $V_{\bot}^y = 140E_R$.
Since we acquire a $y$-axis integrated absorption image,
we do not observe any difference in pumping
even if we lower $V_{\bot}^y$ so that the  
tunneling time along the $y$-direction is comparable to the pumping period.

We also confirm the absence of the shift of the cloud along the direction perpendicular to the pumping direction.
For this purpose, we perform pumping experiments along the $z$-axis and the $x$-axis.
In Fig.~S\ref{fig:reverse} we plot the CoM position along the pumping direction
($z$-axis for \textbf{a} and $x$-axis for \textbf{b}) and the perpendicular direction
($x$-axis for \textbf{a} and $z$-axis for \textbf{b})
for both superlattice configurations (see insets in Fig.~S\ref{fig:reverse}a and b).
As one can see, there is no spatial shift along the perpendicular directions,
which is consistent with our system being an array of one-dimensional superlattices.

To ensure that most of the atoms are loaded into the lowest band of the superlattice,
we first ramp up the long lattice potential in 400~ms, whose band gap is much larger than that of the superlattice. We then ramp up the additional short lattice potential in 150~ms.
We observed that most of the atoms are mapped to the 1st Brillouin zone in our band mapping measurements, which corresponds to about 90\% population in the lowest band.

Moreover, we also confirmed that the atoms are always in the lowest band of the superlattice
during one pumping sequence by using the ``sublattice mapping technique'' developed in our recent Lieb lattice experiment~\cite{Taie15_S}.
In this method, we first quickly change the lattice potential to a ``mapping lattice'' 
with $(V_S, V_{1064}^z)=(40, 10)E_R$ and a fixed relative phase $\phi=0$ in 0.3~ms right after the pumping sequence. 
In this mapping lattice configuration, the A-sublattice $\{a_i\}$ is energetically well separated from the B-sublattice $\{b_i\}$ and the lowest two bands consist of the A- and B-sublattices, respectively.
This maps sublattice occupations after the pumping sequence to band occupations,
which then can be measured by band mapping techniques.
Figure~S\ref{fig:BM}a, b and c show a schematic description of our sublattice mapping technique.
In Fig.~S\ref{fig:BM}d, we show the sublattice mapping measurements
for a cRM pumping sequence with the potential depths of $(V_S, V_L)=(20, 26)E_R$.
As one can see in Fig.~S\ref{fig:BM}d, the band mapping result at $t=0$, $t=T$ and $t=2T$ are almost the same,
namely the atoms in the lowest band are still in the lowest band after pumping sequences.
This result suggests that atoms localized on the B-sublattice $\{b_i\}$ 
can tunnel in the double-wells and then localize on the A-sublattice $\{a_i\}$. The atoms are well described by the lowest-band Wannier functions of the superlattice.
The result also suggests that the atoms are mainly localized at A-sublattice or B-sublattice
in our deep-lattice cRM pumping sequence and that the tunneling only occurs when the Berry curvature has finite value.


\begin{figure*}[bt]
 \includegraphics[width=15cm,clip]{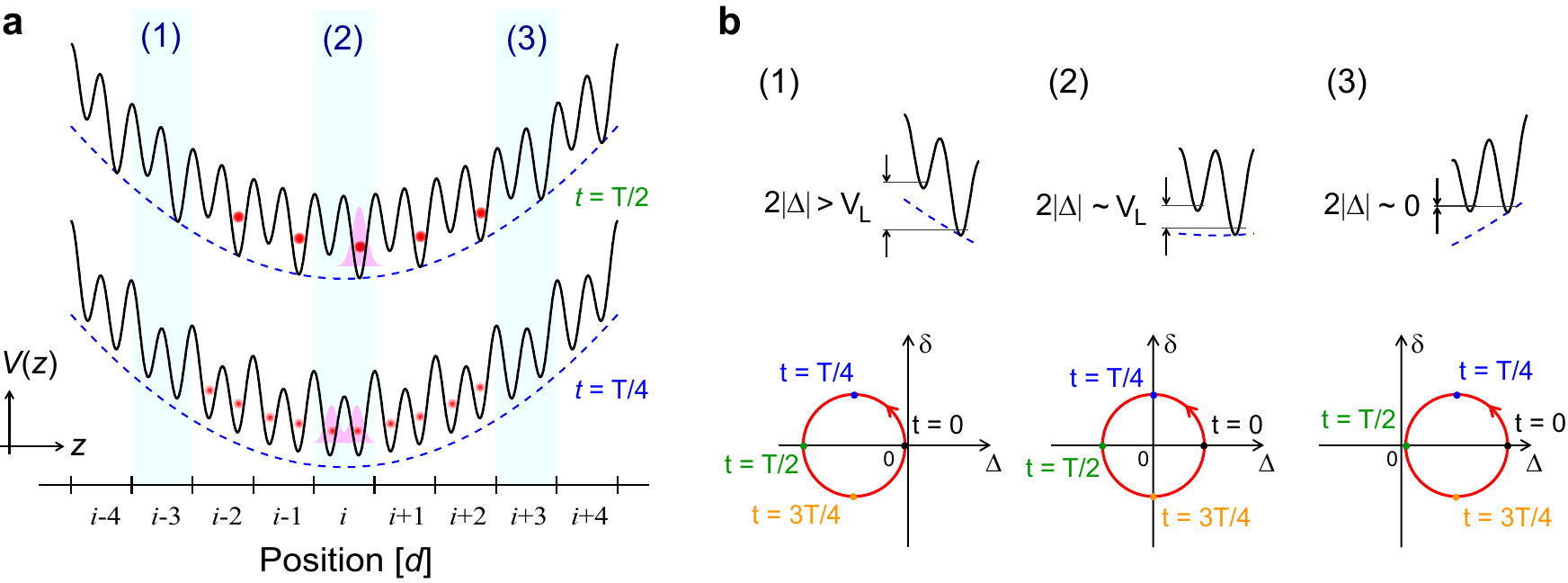}
\caption{
{\bf Saturation of cRM pumping in a harmonic trap.} 
\textbf{a}. Schematics of the Rice-Mele superlattice in harmonic confinement. 
\textbf{b}. The local lattice potentials in three different regions (1)-(3) are shown in the upper part. The lower figures of \textbf{b} are  schematic trajectories of the Rice-Mele pump in these three regions.}
\label{fig:hc_effect}
\end{figure*}

\section*{S5. Details of the \textit{in situ} imaging}
In this section, we describe the details of our \textit{in situ} imaging. 
To evaluate a CoM shift of an atomic cloud in real space,
we measure the CoM position from the \textit{in situ} absorption imaging with a CCD camera (iXon, Andor).
The CoM position is derived from the Gaussian fitting to the integrated one-dimensional data. 
In our imaging setup, one pumping cycle, namely the cloud shift of $d=532$~nm
corresponds to 0.228 pixel on the CCD. The observed cloud width (FWHM) in the pumping direction is typically 3.7(5) pixels ($\sim16(2)d$) except for the higher temperature cases in Fig.~4b.

Note that the observed \textit{in situ} cloud width is not the real atomic cloud width but a convoluted width with the finite optical resolution of our imaging system of $\sim 3~\mu$m ($\sim 6d$).
One notices that the CoM shift after one pumping cycle is less than one pixel,
and much smaller than our imaging resolution.
However, since the Gaussian fit uses many data points (typically more than 20 pixels)
as one can see in Fig.~2c of the main text,
we can reduce the fit uncertainty ($1\sigma$ confidence) down to 0.3$d$ for each image in Fig. 2 of the main text.
Thus the error bars in the CoM measurements mainly come from 
fluctuations of the cloud position, probably due to the fluctuation of the initial atom loading.
Although we cannot suppress this fluctuation, 
we can reduce the statistical error to be smaller than the fitting uncertainty by taking a lot of data.
You can see such an example in Fig.~S\ref{fig:simulation}c, 
in which we evaluate the detailed CoM shift in the small pumping time $0\leq t \leq T$
with more than 30 independent measurements for each data point.

\begin{figure}[tb]
 \includegraphics[width=8cm,clip]{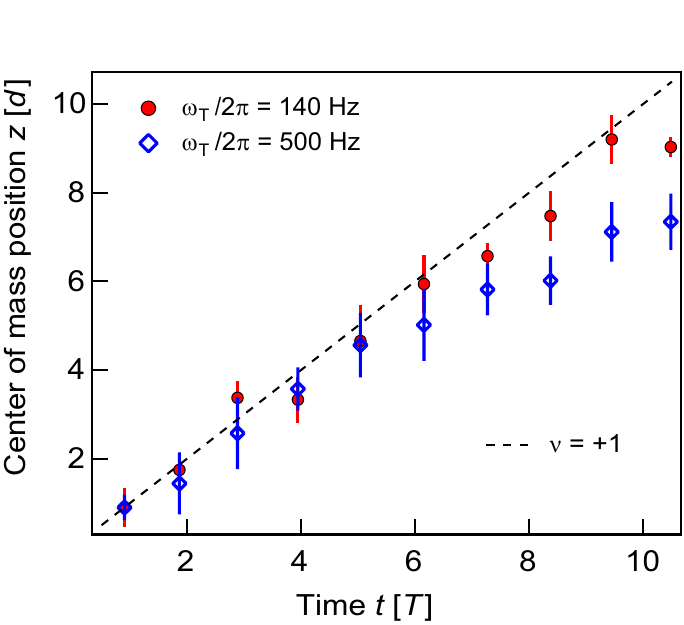}
\caption{
{\bf Saturating behaviour due to the harmonic confinement.} 
The cRM pumping with depths of $(V_S, V_L)=(20, 36)E_R$ at different confinement strengths.
In the case of stronger confinement pumping ($\omega_{T}/2\pi \sim 500$~Hz, blue diamonds)
starts to be saturated after a smaller number of cycles
than that for the case of the weaker confinement ($\omega_{T}/2\pi \sim 140$~Hz, red circles).
}
\label{fig:confinement}
\end{figure}

\section*{S6. Saturation of pumping due to a harmonic confinement}
In this section we discuss a saturation of the pumping due to the harmonic confinement.
As discussed in S2, our system is finite and has the metallic edge states due to the harmonic confinement (Fig.~S\ref{fig:hc_effect}a) while its contribution can be negligible in our deep optical lattice system whose band width is much smaller than the band gap.
As the cloud shifts from the center of the trap, however, a larger position of the cloud will have a local chemical potential inside the lowest energy band, thus being metallic. 
This can also be understood in the terms of the trajectory of the Rice-Mele pump 
as shown in Fig.~S\ref{fig:hc_effect}b. 
The left (right) edge of the trajectory gets close to the degeneracy point as the cloud shifts to right (left), namely, the effective band gap (the minimum distance from the degeneracy point) becomes smaller.
This suggests that the pumping start saturating after a certain CoM shift, depending on the strength of the confinement and the chemical potential or temperature,
while an ideal pumping is possible before this saturation point.
In Fig. 2 (and also in Fig. 3) of the main text, one can indeed see that the pumped amount starts to saturate after several cycles. 
Figure~\figureref{fig:confinement} compares the behavior of the saturation for two different confinement strengths, which clearly shows that larger confinement induces an earlier onset of saturation.

This saturation point is also related to the value of the band gap $D$ through $L_G$.
We actually observed that the saturation point becomes lower as we tune the long lattice depth $V_L$ shallower corresponding to smaller minimum band gap.
These observations suggest that the saturation point is determined by the competition between the band gap and the harmonic confinement.
In other words, we can conclude that the atomic cloud are robustly pumped as long as the pumping system are protected by the gap.


\begin{thebibliography}{10}
\expandafter\ifx\csname url\endcsname\relax
  \def\url#1{\texttt{#1}}\fi
\expandafter\ifx\csname urlprefix\endcsname\relax\def\urlprefix{URL }\fi
\providecommand{\bibinfo}[2]{#2}
\providecommand{\eprint}[2][]{\url{#2}}

\bibitem{Thouless}
\bibinfo{author}{Thouless, D.~J.}
\newblock \bibinfo{title}{Quantization of particle transport}.
\newblock \emph{\bibinfo{journal}{Phys. Rev. B}}
  \textbf{\bibinfo{volume}{27}}, \bibinfo{pages}{6083--6087}
  (\bibinfo{year}{1983}).

\bibitem{RevModPhys.51.591}
\bibinfo{author}{Mermin, N.~D.}
\newblock \bibinfo{title}{The topological theory of defects in ordered media}.
\newblock \emph{\bibinfo{journal}{Rev. Mod. Phys.}}
  \textbf{\bibinfo{volume}{51}}, \bibinfo{pages}{591--648}
  (\bibinfo{year}{1979}).
 
\bibitem{thouless1998topological}
\bibinfo{author}{Thouless, D.~J.}
\newblock \emph{\bibinfo{title}{Topological quantum numbers in nonrelativistic
  physics}} (\bibinfo{publisher}{World Scientific Singapore},
  \bibinfo{year}{1998}).

\bibitem{RevModPhys.82.3045}
\bibinfo{author}{Hasan, M.~Z.} \& \bibinfo{author}{Kane, C.~L.}
\newblock \bibinfo{title}{\textit{Colloquium} : Topological insulators}.
\newblock \emph{\bibinfo{journal}{Rev. Mod. Phys.}}
  \textbf{\bibinfo{volume}{82}}, \bibinfo{pages}{3045--3067}
  (\bibinfo{year}{2010}).

\bibitem{PhysRevLett.45.494}
\bibinfo{author}{Klitzing, K.~v.}, \bibinfo{author}{Dorda, G.} \&
  \bibinfo{author}{Pepper, M.}
\newblock \bibinfo{title}{New method for high-accuracy determination of the
  fine-structure constant based on quantized Hall resistance}.
\newblock \emph{\bibinfo{journal}{Phys. Rev. Lett.}}
  \textbf{\bibinfo{volume}{45}}, \bibinfo{pages}{494--497}
  (\bibinfo{year}{1980}).

\bibitem{TKNN}
\bibinfo{author}{Thouless, D.~J.}, \bibinfo{author}{Kohmoto, M.},
  \bibinfo{author}{Nightingale, M.~P.} \& \bibinfo{author}{den Nijs, M.}
\newblock \bibinfo{title}{Quantized Hall conductance in a two-dimensional
  periodic potential}.
\newblock \emph{\bibinfo{journal}{Phys. Rev. Lett.}}
  \textbf{\bibinfo{volume}{49}}, \bibinfo{pages}{405--408}
  (\bibinfo{year}{1982}).

\bibitem{Altshuler}
\bibinfo{author}{Altshuler, B.~L.} \& \bibinfo{author}{Glazman, L.~I.}
\newblock \bibinfo{title}{Pumping electrons}.
\newblock \emph{\bibinfo{journal}{Science}} \textbf{\bibinfo{volume}{283}},
  \bibinfo{pages}{1864--1865} (\bibinfo{year}{1999}).

\bibitem{Switkes}
\bibinfo{author}{Switkes, M.}, \bibinfo{author}{Marcus, C.~M.},
  \bibinfo{author}{Campman, K.} \& \bibinfo{author}{Gossard, A.~C.}
\newblock \bibinfo{title}{An adiabatic quantum electron pump}.
\newblock \emph{\bibinfo{journal}{Science}} \textbf{\bibinfo{volume}{283}},
  \bibinfo{pages}{1905--1908} (\bibinfo{year}{1999}).

\bibitem{Blumenthal}
\bibinfo{author}{Blumenthal, M.~D.} \emph{et~al.}
\newblock \bibinfo{title}{Gigahertz quantized charge pumping}.
\newblock \emph{\bibinfo{journal}{Nature Phys.}} \textbf{\bibinfo{volume}{3}},
  \bibinfo{pages}{343--347} (\bibinfo{year}{2007}).

\bibitem{Kaestner}
\bibinfo{author}{Kaestner, B.} \emph{et~al.}
\newblock \bibinfo{title}{Single-parameter nonadiabatic quantized charge
  pumping}.
\newblock \emph{\bibinfo{journal}{Phys. Rev. B}}
  \textbf{\bibinfo{volume}{77}}, \bibinfo{pages}{153301}
  (\bibinfo{year}{2008}).

\bibitem{Shilton}
\bibinfo{author}{Shilton, J.~M.} \emph{et~al.}
\newblock \bibinfo{title}{High-frequency single-electron transport in a
  quasi-one-dimensional GaAs channel induced by surface acoustic waves}.
\newblock \emph{\bibinfo{journal}{J. Phys. Condens. Matter}}
  \textbf{\bibinfo{volume}{8}}, \bibinfo{pages}{L531--L539} (\bibinfo{year}{1996}).

\bibitem{Aidelsburger13}
\bibinfo{author}{Aidelsburger, M.} \emph{et~al.}
\newblock \bibinfo{title}{Realization of the Hofstadter Hamiltonian with
  ultracold atoms in optical lattices}.
\newblock \emph{\bibinfo{journal}{Phys. Rev. Lett.}}
  \textbf{\bibinfo{volume}{111}}, \bibinfo{pages}{185301}
  (\bibinfo{year}{2013}).

\bibitem{Miyake13}
\bibinfo{author}{Miyake, H.}, \bibinfo{author}{Siviloglou, G.~A.},
  \bibinfo{author}{Kennedy, C.~J.}, \bibinfo{author}{Burton, W.~C.} \&
  \bibinfo{author}{Ketterle, W.}
\newblock \bibinfo{title}{Realizing the Harper Hamiltonian with laser-assisted
  tunneling in optical lattices}.
\newblock \emph{\bibinfo{journal}{Phys. Rev. Lett.}}
  \textbf{\bibinfo{volume}{111}}, \bibinfo{pages}{185302}
  (\bibinfo{year}{2013}).

\bibitem{Jotzu14}
\bibinfo{author}{Jotzu, G.} \emph{et~al.}
\newblock \bibinfo{title}{Experimental realization of the topological Haldane
  model with ultracold fermions}.
\newblock \emph{\bibinfo{journal}{Nature}} \textbf{\bibinfo{volume}{515}},
  \bibinfo{pages}{237--240} (\bibinfo{year}{2014}).

\bibitem{Aidelsburger}
\bibinfo{author}{Aidelsburger, M.} \emph{et~al.}
\newblock \bibinfo{title}{Measuring the Chern number of Hofstadter bands with
  ultracold bosonic atoms}.
\newblock \emph{\bibinfo{journal}{Nature Phys.}} \textbf{\bibinfo{volume}{11}},
  \bibinfo{pages}{162--166} (\bibinfo{year}{2015}).

\bibitem{Mancini:2015wl}
\bibinfo{author}{Mancini, M.} \emph{et~al.}
\newblock \bibinfo{title}{Observation of chiral edge states with neutral
  fermions in synthetic Hall ribbons}.
\newblock \emph{\bibinfo{journal}{Science}} \textbf{\bibinfo{volume}{349}},
  \bibinfo{pages}{1510--1513} (\bibinfo{year}{2015}).

\bibitem{Stuhl:2015wn}
\bibinfo{author}{Stuhl, B.~K.}, \bibinfo{author}{Lu, H.-I.},
  \bibinfo{author}{Aycock, L.~M.}, \bibinfo{author}{Genkina, D.} \&
  \bibinfo{author}{Spielman, I.~B.}
\newblock \bibinfo{title}{Visualizing edge states with an atomic Bose gas in
  the quantum Hall regime}.
\newblock \emph{\bibinfo{journal}{Science}} \textbf{\bibinfo{volume}{349}},
  \bibinfo{pages}{1514--1518} (\bibinfo{year}{2015}).

\bibitem{Wang}
\bibinfo{author}{Wang, L.}, \bibinfo{author}{Troyer, M.} \&
  \bibinfo{author}{Dai, X.}
\newblock \bibinfo{title}{Topological charge pumping in a one-dimensional
  optical lattice}.
\newblock \emph{\bibinfo{journal}{Phys. Rev. Lett.}}
  \textbf{\bibinfo{volume}{111}}, \bibinfo{pages}{026802}
  (\bibinfo{year}{2013}).

\bibitem{RM}
\bibinfo{author}{Rice, M.~J.} \& \bibinfo{author}{Mele, E.~J.}
\newblock \bibinfo{title}{Elementary excitations of a linearly conjugated
  diatomic polymer}.
\newblock \emph{\bibinfo{journal}{Phys. Rev. Lett.}}
  \textbf{\bibinfo{volume}{49}}, \bibinfo{pages}{1455--1459}
  (\bibinfo{year}{1982}).

\bibitem{Atala}
\bibinfo{author}{Atala, M.} \emph{et~al.}
\newblock \bibinfo{title}{Direct measurement of the Zak phase in topological
  Bloch bands}.
\newblock \emph{\bibinfo{journal}{Nature Phys.}} \textbf{\bibinfo{volume}{9}},
  \bibinfo{pages}{795--800} (\bibinfo{year}{2013}).

\bibitem{Kitagawa08}
\bibinfo{author}{Kitagawa, M.} \emph{et~al.}
\newblock \bibinfo{title}{Two-color photoassociation spectroscopy of ytterbium
  atoms and the precise determinations of s-wave scattering lengths}.
\newblock \emph{\bibinfo{journal}{Phys. Rev. A}}
  \textbf{\bibinfo{volume}{77}}, \bibinfo{pages}{012719}
  (\bibinfo{year}{2008}).

\bibitem{PhysRevA.84.013608}
\bibinfo{author}{Qian, Y.}, \bibinfo{author}{Gong, M.} \&
  \bibinfo{author}{Zhang, C.}
\newblock \bibinfo{title}{Quantum transport of bosonic cold atoms in
  double-well optical lattices}.
\newblock \emph{\bibinfo{journal}{Phys. Rev. A}} \textbf{\bibinfo{volume}{84}},
  \bibinfo{pages}{013608} (\bibinfo{year}{2011}).

\bibitem{Xiao}
\bibinfo{author}{Xiao, D.}, \bibinfo{author}{Chang, M.-C.} \&
  \bibinfo{author}{Niu, Q.}
\newblock \bibinfo{title}{Berry phase effects on electronic properties}.
\newblock \emph{\bibinfo{journal}{Rev. Mod. Phys.}}
  \textbf{\bibinfo{volume}{82}}, \bibinfo{pages}{1959--2007}
  (\bibinfo{year}{2010}).

\bibitem{Shen}
\bibinfo{author}{Shen, S.-Q.}
\newblock \emph{\bibinfo{title}{Topological insulators: Dirac equation in condensed matters}}
(\bibinfo{publisher}{Springer}, \bibinfo{year}{2013}).

\bibitem{Brouwer}
\bibinfo{author}{Brouwer, P.~W.}
\newblock \bibinfo{title}{Scattering approach to parametric pumping}.
\newblock \emph{\bibinfo{journal}{Phys. Rev. B}}
  \textbf{\bibinfo{volume}{58}}, \bibinfo{pages}{R10135--R10138}
  (\bibinfo{year}{1998}).

\bibitem{Mandel}
\bibinfo{author}{Mandel, O.} \emph{et~al.}
\newblock \bibinfo{title}{Coherent transport of neutral atoms in spin-dependent
  optical lattice potentials}.
\newblock \emph{\bibinfo{journal}{Phys. Rev. Lett.}}
  \textbf{\bibinfo{volume}{91}}, \bibinfo{pages}{010407}
  (\bibinfo{year}{2003}).

\bibitem{Fu}
\bibinfo{author}{Fu, L.} \& \bibinfo{author}{Kane, C.~L.}
\newblock \bibinfo{title}{Time reversal polarization and a $Z_2$ adiabatic spin
  pump}.
\newblock \emph{\bibinfo{journal}{Phys. Rev. B}}
  \textbf{\bibinfo{volume}{74}}, \bibinfo{pages}{195312}
  (\bibinfo{year}{2006}).

\bibitem{AA}
\bibinfo{author}{Aubry, S.} \& \bibinfo{author}{Andr{\'e}, G.}
\newblock \bibinfo{title}{Analyticity breaking and Anderson localization in
  incommensurate lattices}.
\newblock \emph{\bibinfo{journal}{Ann. Israel Phys. Soc}}
  \textbf{\bibinfo{volume}{3}}, \bibinfo{pages}{133--140} (\bibinfo{year}{1980}).

\bibitem{Marra}
\bibinfo{author}{Marra, P.}, \bibinfo{author}{Citro, R.} \&
  \bibinfo{author}{Ortix, C.}
\newblock \bibinfo{title}{Fractional quantization of the topological charge
  pumping in a one-dimensional superlattice}.
\newblock \emph{\bibinfo{journal}{Phys. Rev. B}}
  \textbf{\bibinfo{volume}{91}}, \bibinfo{pages}{125411}
  (\bibinfo{year}{2015}).

\bibitem{wei2015anomalous}
\bibinfo{author}{Wei, R.} \& \bibinfo{author}{Mueller, E.~J.}
\newblock \bibinfo{title}{Anomalous charge pumping in a one-dimensional optical
  superlattice}.
\newblock \emph{\bibinfo{journal}{Phys. Rev. A}}
  \textbf{\bibinfo{volume}{92}}, \bibinfo{pages}{013609}
  (\bibinfo{year}{2015}). 

\bibitem{Lohse}
\bibinfo{author}{Lohse, M.} \emph{et~al.}
\newblock \bibinfo{title}{A Thouless quantum pump with ultracold bosonic atoms in an optical superlattice}.
\newblock Preprint at http://arXiv.org/abs/1507.02225 (\bibinfo{year}{2015}).

\bibitem{Taie10}
\bibinfo{author}{Taie, S.} \emph{et~al.}
\newblock \bibinfo{title}{Realization of a SU(2) $\times$ SU(6) system of fermions in
  a cold atomic gas}.
\newblock \emph{\bibinfo{journal}{Phys. Rev. Lett.}}
  \textbf{\bibinfo{volume}{105}}, \bibinfo{pages}{190401}
  (\bibinfo{year}{2010}).

\bibitem{Taie15}
\bibinfo{author}{Taie, S.} \emph{et~al.}
\newblock \bibinfo{title}{Coherent driving and freezing of bosonic matter wave
  in an optical Lieb lattice}.
\newblock \emph{\bibinfo{journal}{Sci. Adv.}}
  \textbf{\bibinfo{volume}{1}},  \bibinfo{pages}{e1500854}
  (\bibinfo{year}{2015}).

\end{thebibliography}

\begin{thebibliography}{10}
\expandafter\ifx\csname url\endcsname\relax
  \def\url#1{\texttt{#1}}\fi
\expandafter\ifx\csname urlprefix\endcsname\relax\def\urlprefix{URL }\fi
\providecommand{\bibinfo}[2]{#2}
\providecommand{\eprint}[2][]{\url{#2}}

\bibitem{Thouless_S}
\bibinfo{author}{Thouless, D.~J.}
\newblock \bibinfo{title}{Quantization of particle transport}.
\newblock \emph{\bibinfo{journal}{Phys. Rev. B}}
  \textbf{\bibinfo{volume}{27}}, \bibinfo{pages}{6083--6087}
  (\bibinfo{year}{1983}).

\bibitem{KS93_S}
\bibinfo{author}{King-Smith, R.~D.} \& \bibinfo{author}{Vanderbilt, D.}
\newblock \bibinfo{title}{Theory of polarization of crystalline solids}.
\newblock \emph{\bibinfo{journal}{Phys. Rev. B}}
  \textbf{\bibinfo{volume}{47}}, \bibinfo{pages}{1651--1654}
  (\bibinfo{year}{1993}).

\bibitem{Xiao_S}
\bibinfo{author}{Xiao, D.}, \bibinfo{author}{Chang, M.-C.} \&
  \bibinfo{author}{Niu, Q.}
\newblock \bibinfo{title}{Berry phase effects on electronic properties}.
\newblock \emph{\bibinfo{journal}{Rev. Mod. Phys.}}
  \textbf{\bibinfo{volume}{82}}, \bibinfo{pages}{1959--2007}
  (\bibinfo{year}{2010}).

\bibitem{Bernevig_S}
\bibinfo{author}{Bernevig, B.~A.} \& \bibinfo{author}{Hughes, T.~L.}
\newblock \emph{\bibinfo{title}{Topological insulators and topological
  superconductors}} (\bibinfo{publisher}{Princeton University Press},
  \bibinfo{year}{2013}).

\bibitem{Zak_S}
\bibinfo{author}{Zak, J.}
\newblock \bibinfo{title}{Berry's phase for energy bands in solids}.
\newblock \emph{\bibinfo{journal}{Phys. Rev. Lett.}}
  \textbf{\bibinfo{volume}{62}}, \bibinfo{pages}{2747--2750}
  (\bibinfo{year}{1989}).

\bibitem{TKNN_S}
\bibinfo{author}{Thouless, D.~J.}, \bibinfo{author}{Kohmoto, M.},
  \bibinfo{author}{Nightingale, M.~P.} \& \bibinfo{author}{den Nijs, M.}
\newblock \bibinfo{title}{Quantized hall conductance in a two-dimensional
  periodic potential}.
\newblock \emph{\bibinfo{journal}{Phys. Rev. Lett.}}
  \textbf{\bibinfo{volume}{49}}, \bibinfo{pages}{405--408}
  (\bibinfo{year}{1982}).

\bibitem{Aidelsburger_S}
\bibinfo{author}{Aidelsburger, M.} \emph{et~al.}
\newblock \bibinfo{title}{Measuring the chern number of hofstadter bands with
  ultracold bosonic atoms}.
\newblock \emph{\bibinfo{journal}{Nat Phys}} \textbf{\bibinfo{volume}{11}},
  \bibinfo{pages}{162--166} (\bibinfo{year}{2015}).

\bibitem{Wang_S}
\bibinfo{author}{Wang, L.}, \bibinfo{author}{Troyer, M.} \&
  \bibinfo{author}{Dai, X.}
\newblock \bibinfo{title}{Topological charge pumping in a one-dimensional
  optical lattice}.
\newblock \emph{\bibinfo{journal}{Phys. Rev. Lett.}}
  \textbf{\bibinfo{volume}{111}}, \bibinfo{pages}{026802}
  (\bibinfo{year}{2013}).

\bibitem{Kitagawa:2010bu}
\bibinfo{author}{Kitagawa, T.}, \bibinfo{author}{Berg, E.},
  \bibinfo{author}{Rudner, M.} \& \bibinfo{author}{Demler, E.}
\newblock \bibinfo{title}{{Topological characterization of periodically driven
  quantum systems}}.
\newblock \emph{\bibinfo{journal}{Phys. Rev. B}} \textbf{\bibinfo{volume}{82}},
  \bibinfo{pages}{235114} (\bibinfo{year}{2010}).

\bibitem{RM_S}
\bibinfo{author}{Rice, M.~J.} \& \bibinfo{author}{Mele, E.~J.}
\newblock \bibinfo{title}{Elementary excitations of a linearly conjugated
  diatomic polymer}.
\newblock \emph{\bibinfo{journal}{Phys. Rev. Lett.}}
  \textbf{\bibinfo{volume}{49}}, \bibinfo{pages}{1455--1459}
  (\bibinfo{year}{1982}).

\bibitem{Taie15_S}
\bibinfo{author}{Taie, S.} \emph{et~al.}
\newblock \bibinfo{title}{Coherent driving and freezing of bosonic matter wave
  in an optical Lieb lattice}.
\newblock \emph{\bibinfo{journal}{Sci. Adv.}}
  \textbf{\bibinfo{volume}{1}},  \bibinfo{pages}{e1500854}
  (\bibinfo{year}{2015}).
  
\end{thebibliography}

\providecommand{\noopsort}[1]{}\providecommand{\singleletter}[1]{#1}%

\end{document}